\documentclass[structabstract]{aa}
\usepackage{graphicx}
\usepackage{txfonts}
\usepackage{verbatim}
\usepackage{natbib}
\bibpunct{(}{)}{;}{a}{}{,} 

\begin{document}

\title{Flip-flops of \object{FK Comae Berenices}\thanks{Based on data obtained 
with the Amadeus T7 Automatic Photoelectric Telescope (APT)
at Fairborn Observatory, 
jointly operated by the University of Vienna and AIP, 
the Phoenix-10 APT at Mt.~Hopkins, 
Arizona, and the Nordic Optical Telescope,
Observatorio Roque de los Muchachos, La Palma, Canary Islands. The photometric
observations are available in electronic form at the CDS via
anonymous ftp to cdsarc.u-strasbg.fr (130.79.128.5) 
or via http://cdsarc.u-strasbg.fr/vizbin/qcat?J/A+A/}}

   \author{T. Hackman \inst{1,} \inst{2} \and
           J. Pelt  \inst{3} \and
           M.J. Mantere \inst{1,} \inst{4} \and
           L. Jetsu \inst{1} \and
           H. Korhonen \inst{5,} \inst{2,} \inst{6} \and
           T. Granzer \inst{7} \and
           P. Kajatkari \inst{1} \and
           J. Lehtinen \inst{1,} \inst{8} \and
           K.G. Strassmeier \inst{7} }

   \offprints{T. Hackman\\
          \email{thomas.hackman@helsinki.fi}
          }   

\institute{Department of Physics, PO Box 64 , FI-00014
              University of Helsinki, Finland
\and
Finnish Centre for Astronomy with ESO (FINCA), University of Turku, 
V\"{a}is\"{a}l\"{a}ntie 20, FI-21500 Piikki\"{o}, Finland
\and 
Tartu Observatory, T\~{o}ravere, 61602, Estonia
\and
Aalto University, Department of Information and Computer Science, 
PO Box 15400, FI-00076 Aalto, Finland
\and
Niels Bohr Institute, University of Copenhagen, Juliane Maries Vej 30, DK-2100 
K{\o}benhavn {\O}, Denmark
\and 
Centre for Star and Planet Formation, Natural History Museum of Denmark, 
University of Copenhagen, {\O}ster Voldgade 5-7, 
DK-1350 K{\o}benhavn {\O}, Denmark
\and
Leibniz-Institut f\"{u}r Astrophysik Potsdam, An der Sternwarte 16, 14882 
Potsdam, Germany
\and
Nordic Optical Telescope, 38700 Santa Cruz de la Palma, Spain
}

   \date{Received ; accepted }

 \abstract{\object{FK Comae} is a rapidly rotating magnetically active star, 
the light curve of which is modulated by cool spots on its surface. It 
was the first star where the ``flip-flop'' phenomenon was discovered.
Since then, flip-flops in the spot activity have been reported in many
other stars. Follow-up studies with increasing length have shown, however, 
that the phenomenon is more complex than was thought right after its 
discovery.}
{Therefore, it is of interest to perform a more thorough study of the
  evolution of the spot activity in FK Com. 
In this study, we analyse 15
years of photometric observations with
  two different time series analysis methods, with a special emphasis on
  detecting flip-flop type events from the data.}
{We apply the continuous period search and carrier fit methods 
on long-term standard Johnson-Cousins V-observations from the years 1995--2010.
The observations were carried out with two automated photometric
telescopes, Phoenix-10 and Amadeus T7 located in Arizona.
}
{We identify complex phase behaviour in 6 of the 15 analysed data
segments. We identify five
flip-flop events and two cases of phase jumps, where the phase shift is 
$\Delta \phi < 0.4$. In addition we see two mergers of spot regions
and two cases where
the apparent phase shifts are caused by spot regions 
drifting with respect to each other.
Furthermore we detect variations in the rotation period 
corresponding to a differential rotation coefficient of $|k|>0.031$.}
{The flip-flop cannot be interpreted as a single phenomenon, where the 
main activity jumps from one active longitude to another. In some of 
our cases the phase shifts can be explained by differential rotation: 
Two spot regions move with different angular velocity and even 
pass each other. 
Comparison between the methods show that the carrier fit utility is better in 
retrieving slow evolution especially from a low amplitude light curve, while 
the continuous period search is more sensitive in case of rapid changes.}
{}


   \maketitle
%

\section{Introduction}

\object{FK Comae Berenices} (\object{HD 117555}; hereafter \object{FK Com}) is
the prototype of a class of single chromospherically active rapidly
rotating G-K giants. Only a few stars fulfil the definition of the
\object{FK Com} class \citep{bopp}. These stars may represent an
intermediate state of coalesced W UMa binaries \citep[e.g.][]{boppsten} 
in the process of magnetic braking, which would explain why they are so rare.

\object{FK Com} itself is an extremely active late-type star and has been 
extensively studied with ground-based optical photometry and spectroscopy, 
radio observations, as well as satellite-based UV- and X-ray observations 
\citep[e.g.][]{jetsu1994a,olah2006,PanovDimitrov,korhonen2009,hughes87,boppsten,ayres06,drake08}.

The photometric rotation period 
of \object{FK Com} is $P_\mathrm{phot} \approx 2\fd4$ \citep{chuga,jetsu1993}.
It has been proposed that its spectral class is between G5 {\sc iii}
\citep{korhonen1999,korhonen2007} and G4 {\sc iii}
\citep{klasu09}.
\citet{korhonen2000} concluded that $v \sin i = 159$ km s$^{-1}$ 
gave the best fit for the spectral data.

Analysing photometry spanning roughly over 25 years (1966--1990),
\citet{jetsu1993} reported a switch of activity between two
longitudes separated by approximately 180\degr, and labelled this
effect the ``flip-flop''. 
The activity was observed to jump from one active longitude to the
other three times during the period of analysis. Furthermore, the
active longitude system was reported to be rotating with the
photometric rotation period of $P_\mathrm{phot}=2.^{\mathrm d}4002466\pm0.^\mathrm{d}0000056$ throughout the whole span of the
data.

The flip-flop phenomenon has since then
been suggested to occur in a number of stars
\citep[e.g.][]{jetsu1996,BT1998,lehtinen2011}. 
While further results of the stable active longitude system producing
flip-flops on \object{FK Com} were published by \citet{jetsu1994b},
evidence for the phenomenon being more complex started building up,
e.g. by the analysis of photometry by \citet{klasu97}. In this
study, a gradual drift of the spots from one active longitude to
the other was detected during 1993--1995, in contrast to
the abrupt changes reported earlier, where spot migration over phase was
not related to the phenomenon. With improved photometric data with
denser timing piling up,
the picture of a steady active longitude system rotating with one
single period was abandoned. For instance in the study of
\citet{korhonen2002}, where one-dimensional photometric inversions of
\object{FK Com} were presented, the active longitude system exhibited
three different periods: in the beginning of the observations, the
system was rotating with the photometric rotation period of the star,
slowed down during 1994--1997, and sped up to super-rotation for
1998--2004 \citep{korhonen2004}. The first apparent semi-regularity of
the flip-flops also became under doubt; the time 
between the flip-flops could range from a year to several years. 
Cycles of 5.2 and 5.8 years in the migration of the two active
longitudes were reported by \citet{olah2006}. Similarly, 
\citet{PanovDimitrov} reported oscillatory spot migration with a cycle length
of $5.8\pm 0.1$ years. 
Furthermore \citet{olah2006} made a
distinction between phase jumps and flip-flops.
In the case of phase jumps, a new active region appears on the same 
hemisphere of the star as the
old active region, resulting in a phase shift of less than 0.5.
During a flip-flop the active longitude changes $\sim 180 \degr$.

The behaviour of the active longitude system and the
related flip-flops were spectroscopically confirmed by analysing the
uniquely long series of Doppler images from the years 1993--2008
by Korhonen and collaborators \citep{korhonen2000,korhonen2007, ayres06, 
korhonen2009, korhonen2009b}.
This work was based mainly on the observations with the SOFIN
high-resolution spectrograph at the Nordic Optical Telescope (La Palma, Spain).

\cite{jetsu1994b} detected variations in the photometric period, 
which could be a signature of differential rotation in \object{FK Com}.
\cite{korhonen2007} estimated the differential rotation by combining Doppler 
images and period analysis of photometry. They reported a rotation law of

\begin{equation}
\Omega \approx (151.30^{\mathrm o}/{\mathrm d} \pm 0.09°/{\mathrm d})-
(1.78^{\mathrm o}/{\mathrm d} \pm 0.12^{\mathrm o}/{\mathrm d}) \sin^2 \psi,
\label{dr}
\end{equation}
\noindent

\noindent where $\psi$ is the stellar
latitude. The estimated relative differential 
rotation coefficient was thus $k \approx 
0.012$.
It should, however, be emphasised
that estimating differential rotation using Doppler imaging
is challenging, especially because of artifacts and errors in the spot 
latitudes.

Observations of \object{FK Com} obtained with the Far Ultraviolet 
Spectroscopic Explorer (FUSE) revealed complex profiles suggesting that the 
transition region and the corona are highly structured, dominated by dynamic 
processes, and could be very extended \citep{ayres06}.
The X-ray observations obtained with XMM-Newton imply that the corona of 
FK Com is dominated by large magnetic structures similar to the
interconnecting loops in solar active regions, but significantly
hotter \citep{gondoin2002}.
Similarly, using data from the Chandra X-ray
Observatory, \cite{drake08} found indications of magnetic loops in the corona 
of FK Com. Their 
observations suggested that the observed X-ray emission originates
from plasma residing predominantly in extended structures centred at a phase 
halfway between two spot regions, and that the coronal structures revealed 
by the Chandra observations correspond to magnetic loops joining these two 
spot regions. 
This would support a model where the two regions have opposite 
magnetic polarities. 

We want to investigate the spot activity in more detail and combining different 
methods. Our main aim is to identify and study the flip-flops of 
\object{FK Com}. We apply two novel time series analysis methods on long-term
photometric observations; the continuous period search
method
\citep[][hereafter CPS]{lehtinen2011} and the carrier fit utility
\citep[][hereafter CF]{pelt11}. In particular, we are interested in
the nature of the flip-flops. 
We want to isolate different types of proposed effects:
i) abrupt
jumps from one hemisphere to another representing the 
flip-flop phenomenon,
ii) events better described as phase jumps (with a phase separation 
significantly less than 180$\degr$), and iii) gradual phase
drifts of the active longitudes. We also aim at establishing whether,
when properly classified and separated, there are any regularities
related to the phenomena.

Our secondary aim is to compare the results from our two analysis
methods. We anticipate that there is a great advantage of combining
the two time series analysis methods because of their different  solutions 
for modelling the data.
It is also important to make a comparative study with real data.


\begin{figure*}
\centering
\vspace{-5.0cm}
 \includegraphics[angle=90,width=18cm]{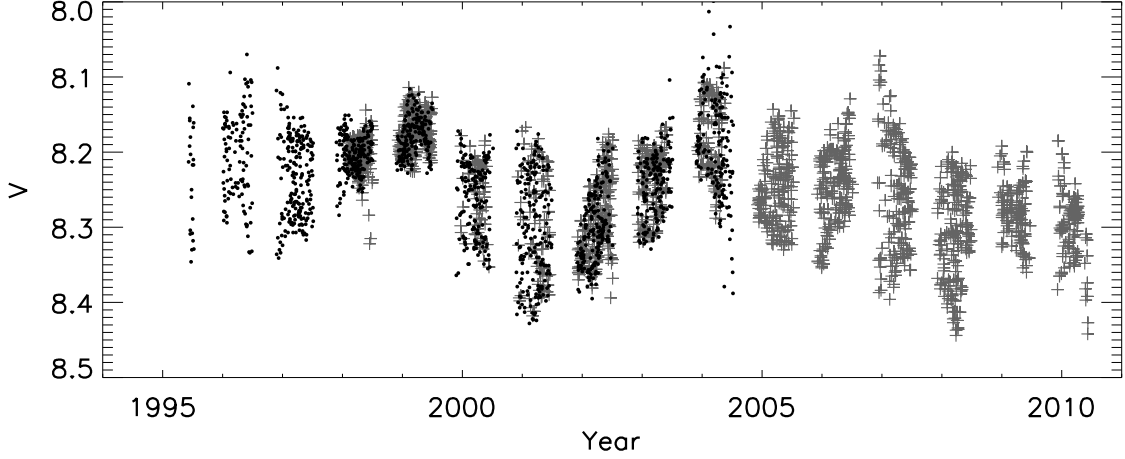}
\vspace{-1.7cm}
      \caption{All photometric V-data for FK Com. The Phoenix-10 data is
marked with black dots and the Amadeus data is marked with 
grey pluses.}
         \label{alldata}
   \end{figure*}

\begin{figure*}
\centering
\vspace{-7.5cm}
 \includegraphics[angle=90,width=18cm]{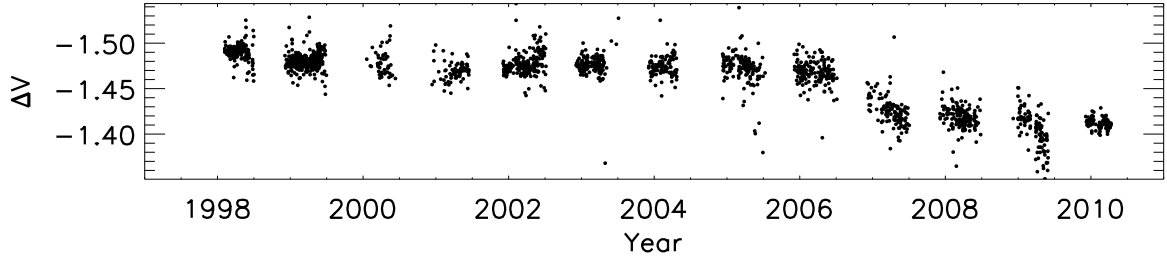}
\vspace{-1.7cm}
      \caption{The differential magnitudes of the comparison star 
\object{HD 117567} for the Amadeus data: 
$\Delta V = V_\mathrm{HD117876} - V_\mathrm{HD117567}$.}
\label{compdata}
   \end{figure*}

\section{Observations} \label{observations}

The photometric standard Johnson-Cousins V observations were collected with the
0.75m Vienna University/AIP APT ``Amadeus'' \citep[][T7 in 
Table\ref{obs}]{amadeus}, located at Fairborn
Observatory and the Phoenix-10 APT at Mt. Hopkins, Arizona (Ph10). Data
points with errors greater than 0.02 were rejected and the
differential magnitudes were transferred into apparent magnitudes using
the comparison star HD 117567 just as in \cite{jetsu1994a}. 
A change corresponding to $\Delta V \sim 0.06$ can be seen in the
difference between the magnitudes of the check star \object{HD 117876} and 
comparison star \object{HD 117567} during 2006 (Fig. \ref{compdata}). The 
spectral type of the 
comparison star is F8, while the check star is classified as G8 {\sc ii-iii}.
Since no corresponding dip can be seen in the observations of \object{FK Com}, 
we concluded that it is the check star which has changed.
For a more detailed description of the observations of FK Comae, we refer 
to the paper by \citet{korhonen2001}.

Most of the data has  been
included in earlier papers 
\citep[][and references therein]{korhonen2002,olah2006,olah2009,korhonen2009}.
The data is summarised in Table~\ref{obs} and published electronically in
the CDS. Note that each observing season forms a segment.
The new data consists of segments SEG14--SEG16.

\begin{table}
\caption{Summary of the observations and labelling of each segment in the text.}
\begin{center}
\begin{tabular}{llll}
\hline \hline
Segment &$t_\mathrm{min}$  &$t_\mathrm{max}$ & Telescope\\
        & HJD-2400000 & HJD-2400000& \\ \hline
SEG1    &49876.7555   &49908.6573 & Ph10 \\
SEG2    &50085.0551   &50265.6741 & Ph10\\
SEG3    &50412.0383   &50636.6781 & Ph10 \\
SEG4    &50778.0423   &50997.7057 & Ph10 \& T7\\
SEG5    &51144.0394   &51362.7280 & Ph10 \& T7\\
SEG6    &51508.0427   &51731.6859 & Ph10 \& T7\\
SEG7    &51873.0421   &52089.7274 & Ph10 \& T7\\
SEG8    &52242.0328   &52461.6984 & Ph10 \& T7\\
SEG9    &52613.0093   &52828.6937 & Ph10 \& T7\\
SEG10   &52972.0288   &53194.6817 & Ph10 \& T7\\
SEG11   &53343.0462   &53565.6808 & T7 \\
SEG12   &53709.0464   &53922.6978 & T7 \\
SEG13   &54075.0453   &54285.6861 & T7\\
SEG14   &54440.0490   &54643.7298 & T7\\
SEG15   &54807.0399   &55004.7444 & T7\\
SEG16   &55172.0443   &55297.8314 & T7\\ \hline
\end{tabular}
\label{obs}
\end{center}
\end{table}

\section{Analysis methods}

Both the CPS and CF analysis methods are based on an approach
of continuous curve fitting. A main difference is that the CPS method allows
the period to vary, while the CF uses a constant carrier
period. Both approaches can be argued for. On one hand e.g. differential
rotation will cause the photometric period to depend on the spot
latitude or the anchoring depth of the spot structure. On the other
hand, noise in the data and insufficient phase coverage will
contribute to spurious period variations in the CPS method
\citep{lehtinen2011}. The CF method is more stable against such
errors.

The main advantage of the CF- and CPS-methods is their 
flexible
approach. The most common time series analysis methods
used in astronomy are based on power 
spectrum analysis, e.g. the ones presented by \cite{deeming}, 
\cite{Sca82} and \cite{Hor86}. These are limited in that there will be 
problems in dealing with higher harmonics and changes in the mean magnitude, 
light curve amplitude and minimum phase. This also means that the advantages 
of the CF- and CPS-methods become important when analysing long series of 
observations with a dense time coverage of a star with changing spot activity. 
The demand of dense timing of the observations is the reason why we omitted 
some part of the observations listed in Section \ref{observations} and 
some previously published
observations of \object{FK Com}, e.g. the observations presented by 
\cite{PanovDimitrov}.

An alternative to time series analysis methods is offered by light curve 
modelling. With the availability of satellite observations, significant 
advances have been made in this area \citep[see e.g.][]{croll06}. However,
light curve modelling involves assumptions about considerable stability of the 
spots, which
is clearly not the case for \object{FK Com} (see e.g. Fig. \ref{seg4-5cffit}).

\subsection{CF method}\label{CF}

The CF method is based on the simple idea of decomposing the
observed stellar light curves into two components: 1) a rapidly
changing carrier modulation tracing the regular part
of the signal, for instance rotation of a spotted star, and 2) a slowly changing
modulation, such as evolution of the cool
spots on the stellar
surface. Such a situation can be described with the following model

\begin{equation}
y_\mathrm{cf}(t) =  a_0(t)  + \sum\limits_{k = 1}^K {\big ( a_k (t) \cos (2\pi kf_0t ) + b_k (t)\sin (2\pi kf _0t )\big )}, 
\label{fullCF}
\end{equation}

\noindent where $a_0(t)$ is the time-dependent mean level of the signal, $K$ is
the total number of harmonics included in the model, describing the
overtones of the basic carrier frequency, while $a_k(t)$ and $b_k(t)$
are the low-frequency signal components. The carrier frequency
$f_0$ can be either known a priori, or determined using the CF utility, as the
first step of the analysis. In this paper, we take the previous
determinations of the photometric rotation period as the first guess
of the carrier period 
$P_0=1/f_0=2^{\mathrm d}.40$. 
The next step in the
analysis is to formulate a suitable model for the modulating
curves. In \citet{pelt11} we introduced two classes of models based on
either trigonometric or spline approximation. In this paper, models
based on the trigonometric approximation are used. 

The trigonometric approximation model for the modulating signals is
built in the following way. Let the time interval $[t_{\min},t_{\max}]$
be the full span of our input data. Then we can introduce a certain
period $D = C\times (t_{\max}-t_{\min})$ for which the coverage factor
$C$ is larger than unity (typically $C=1.1-1.5$). Using the
corresponding frequency, $f_{D}= 1/D$, we can now build a
trigonometric (truncated) series of the type:
\begin{equation}
a(t) = c_0^a  + \sum\limits_{l = 1}^L {\big ( c_l^a\cos (2\pi tlf_D ) + s_l^a\sin (2\pi tlf_D )\big ),}
\end{equation}
and
\begin{equation}
b(t) = c_0^b  + \sum\limits_{l = 1}^L {\big ( c_l^b\cos (2\pi tlf_D ) + s_l^b\sin (2\pi tlf_D )\big ),} 
\end{equation}
where $L$ is the total number of harmonics used in the modulator
model. According to our definition of a slow process, the period $D$
must be significantly longer than the carrier period $P_0$. With 
the data segment lengths in the regime of 100--200 days, this condition is
well satisfied with the chosen coverage factor. Next, proper expansion
coefficient estimates are computed for every term in the series for
the fixed carrier frequency $f_0$ and the ``data frequency'' $f_D$;
this is a standard linear estimation procedure and can be
implemented using standard mathematical (statistical) packages, as
described in detail by \citet{pelt11}. If the coefficients ($a_k,b_k$) consist
of the same number of harmonics $L$ and we approximate separate cycles by a
$K$-harmonic model, then the overall count of linear parameters to be
fitted is $N=(2 \times L+1)*(2 \times K+1)$. The actual choice of the
representative parameters $K$ and $L$ depends on the particular object
we are working with. The number of tones, $K$, depends on the
complexity of the phase curves. The choice of $L$ is constrained by
the longest gaps in the time series. In this study, we adopt $K=2$ and
$L=3$, resulting in the total number of free parameters to be fitted
$N=35$.

We visualise our results in the following way.  First we calculate a
continuous curve least-squares estimate $\hat y_\mathrm{cf}(t)$,
from the randomly spaced and gapped data set.
This approximation is continuous and does not contain gaps,
and therefore allows us to get a smooth picture of the long-term
behaviour. Next we divide this continuous curve into 
strips with a length of the carrier period $P_0 = {1 \over f_0}$. We then
normalise each strip so that the approximating values span the
standard range of $[-1,1]$. After normalisation, we stack the strips
along the time axis. To enhance the obtained plot, we extend
every strip somewhat along phases, so that the actual display is wider (along
phases) than a single period. The normalisation is a relevant part of
our procedure because it helps to grasp the phase information we are
interested in (trends, drifts, flip-flops etc.). This method of
visualisation allows to verify that the model fits into the data and
not into the gaps. 
If the number of nodes or harmonics $L$ for the
modulation model curves is chosen 
properly, then the phase plots do not reveal
any underlying timing structure.

The data from both telescopes were merged for the CF-analysis. In principle 
merging V-magnitudes from two different sources may pose problems, even though 
the reduction is carefully done. In this case however, we concluded
that the data was uniform enough for the CF-analysis. The CF analysis was
applied on segments SEG2--16; in SEG1 there
were not enough points to carry out the 
analysis.

The goodness of the fit was estimated with the coefficient of determination

\begin{equation}
R^2=1-{\sum \limits_i ({y_i - f_i})^2 / \sum \limits_i (y_i - {\bar y})^2},
\label{detcoef}
\end{equation}

\noindent where $\bar y$ is the mean of the data. Our goal was to have
$R^2 > 0.9$. In five out of all the fifteen segments $R^2$ was too small.
Therefore, for SEG4, SEG5, SEG9, SEG10 and SEG15, we refined our 
analysis by excluding the  $3 \sigma$ outliers after an initial
CF fit, and making a new fit.
It is probably no coincidence, that four of these segments consisted of data
from the two different telescopes. However, tests showed that excluding
the outliers had no other significant influence on the result than 
that of improved goodness.

   \begin{figure*}
   \centering
\includegraphics[angle=0,width=\textwidth]{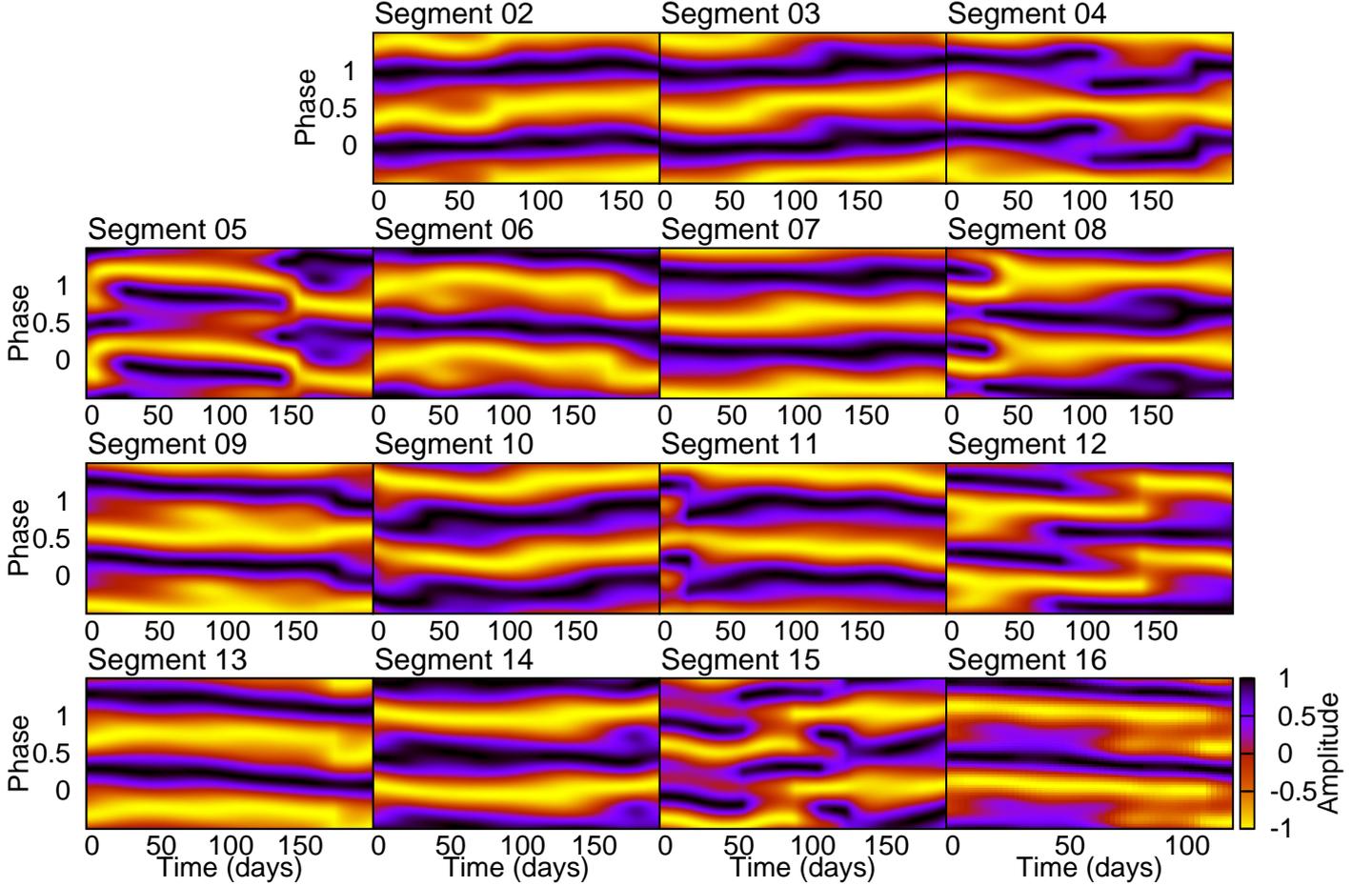}
      \caption{CF analysis results for each segment with the carrier period 
$P_0=2^\mathrm{d}.40$. Each panel shows the computed time-dependent phase 
diagram, i.e. the normalised light curve amplitude
profile over phase (y-axis) plotted as function of time (x-axis).}
\label{alldataCF}
\end{figure*}

\subsection{CPS method}

The CPS method was originally developed for the analysis of photometric 
observations of late-type stars. In 
order to understand the spot activity of these stars, one has to take
into account both short time scale (days) and long-term (years) 
changes \citep{lehtinen2011}. The method is 
based on the Three Stage Period Analysis 
\citep[TSPA,][]{TSPA}. The model

\begin{equation}
y_\mathrm{cps}(t_i,\bar{\beta}) = a_0 + 
\sum_{k=1}^K{[a_k\cos{(2\pi k ft_i)} + b_k\sin{(2\pi k ft_i)}]},
\label{cpsmodel}
\end{equation}

\noindent is used to fit each set of the data. 
Here $\bar{\beta}$ consists of
the parameters $(a_i,b_i,f)$, which are determined through a non-linear least 
squares optimisation.
The three major improvements compared with the TSPA are:

\begin{enumerate}
\item We analyse the data with a sliding window of length $\Delta T_{\max}$  in 
order to increase the time resolution.
\item We test models of different orders and choose the final order $K$ using
a Bayesian information criterion . 
\item We derive the time scale $T_\mathrm{C}$ of significant changes in the 
light curve.
\end{enumerate}

With the CPS-analysis one can thus derive a continuous series of 
estimates for the 
mean magnitude ($M$), total light curve amplitude ($A$), photometric 
period ($P$) and epochs of the light curve minima ($t_{\min}$).
The temporal changes of the mean and amplitude are useful for studying 
variations in the level of spot activity, since the mean magnitude will be
sensitive to the spot coverage and the amplitude is a measure of the 
non-axisymmetry of the spot configuration. 
Variations in the photometric period can be caused by differential rotation, 
or alternatively dynamo waves \citep{krause1980,tuominen2002}. The time scale
of significant change $T_\mathrm{C}$ can be used to estimate the stability of
the light curve. It may also be related to the convective turnover time 
$\tau_\mathrm{c}$ \citep{lehtinen2011,lehtinen2012}.

The CPS-analysis was applied to the Phoenix 10 and Amadeus data separately. 
This method uses less data points in each fit than the CF-method, which makes
it more vulnerable to errors in the data. Furthermore, with $K=2$ 
the number of free parameters is six, i.e. much less than in the CF-method. 
Thus, the need to maximise the number of data points is less important
than the homogeneity of the data for the CPS analysis.  The maximum
length of the moving window was $\Delta T_{\max}=24$d. This length was chosen
because the rotation period being $\sim 2\fd4$, it will give an
optimal phase coverage in the case of evenly spaced observations.
Normally $\Delta T_{\max}$ would also define the division of the data into
segments, as described in \citet{lehtinen2011}. However, for
consistency reasons in this paper
the segments are identical to the observing seasons.
The maximum order of the fit, i.e. the largest tested $K$-value for the 
model (Eq. \ref{cpsmodel}), was $K_\mathrm{lim}=2$. 

\begin{figure*}
\centering
\includegraphics[trim=1cm 2.2cm 0cm 1.5cm, clip=true, angle=-90,width=9cm]{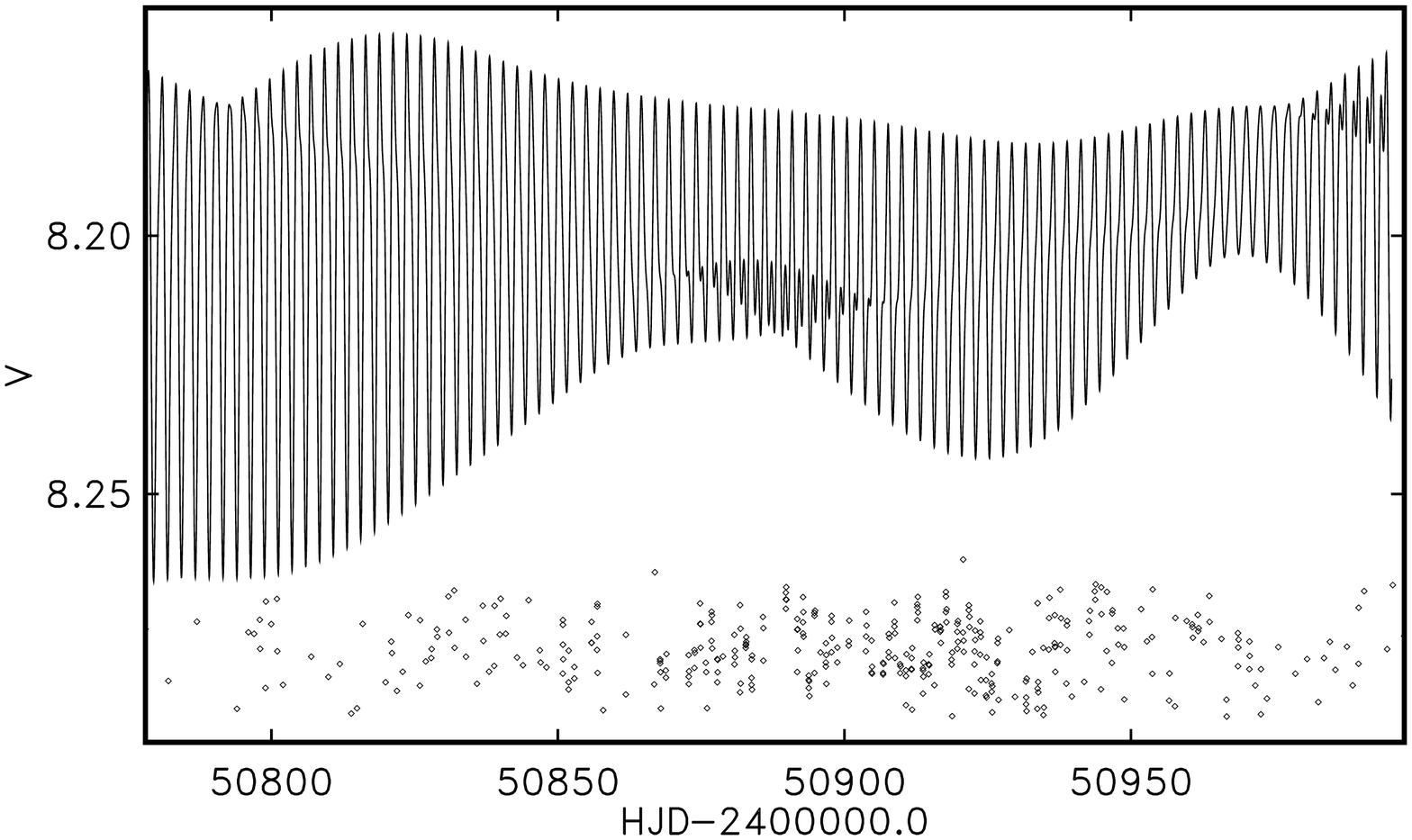}
\includegraphics[trim=1cm 2.2cm 0cm 1.5cm, clip=true, angle=-90,width=9cm]{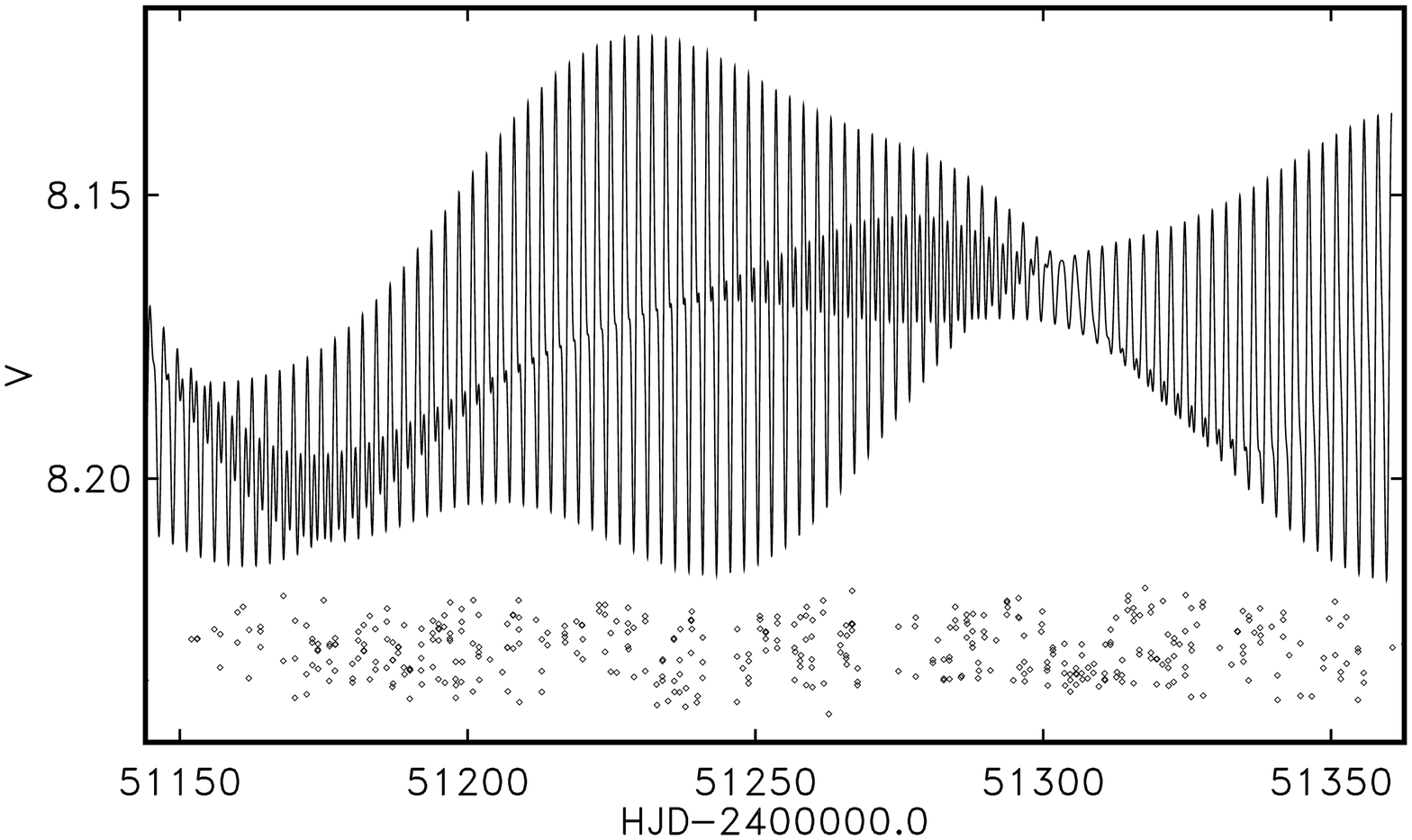}
\caption{The CF fits for segments SEG4 and SEG5 
using the carrier period $P_0=2\fd40$.
The residuals at the bottom of
the plot are shifted with 8.28 (SEG4) and 8.23 (SEG5).}
\label{seg4-5cffit}
\end{figure*}

\begin{figure*}
\vspace{1.5cm}
\centering
\includegraphics[angle=90,width=15cm]{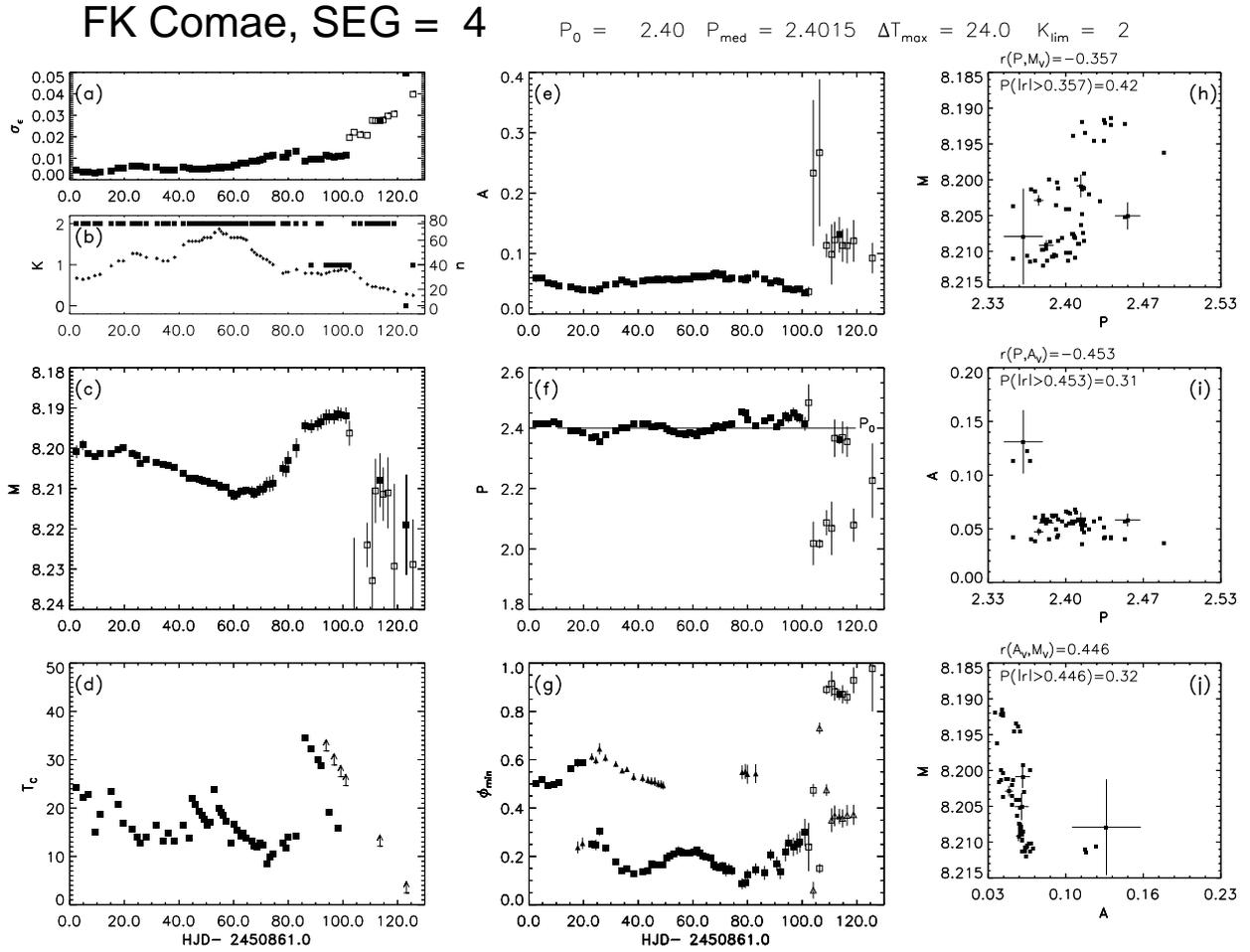}
\caption{The CPS analysis of segment SEG4 of the 
Amadeus photometry of \object{FK Com}. The phases in panel (g) where 
calculated using the period $P_\mathrm{med}=2\fd4015$. Further descriptions 
of the subplots are given in Sect. \ref{results}.}
\label{seg4}
\end{figure*}
 
\subsection{Kuiper test}

Non-parametrical time-series analysis methods, such as the Kuiper test
for phase distributions
\citep{kuiper1960}, can be utilised to identify active longitudes from the
epochs of light curve minima.
We used the unweighted Kuiper-test as formulated by \citet{JP1996}. The
Kuiper periodogram is calculated for a set of epochs of photometric
minima $t_{\min}$. The most significant periods are tested against the null 
hypothesis of a random phase distribution. Examples of
application of this method can be found in \citet{jetsu1996},
\citet{lehtinen2011} and \citet{lehtinen2012}.

We computed the Kuiper statistic periodogram for periodicities of 
2.2 -- 2.6 days  using the epochs of photometric minima $t_{\min}$ derived
both with the CF and CPS methods. 
A total of 1637 primary and secondary minima where retrieved from the 
CF-analysis. From the CPS-analysis we used separately the 136 independent 
minima and all 1837 minima.
In addition to the minima, we also tested 1640 maxima from 
the CF-analysis.
%


\begin{figure*}
\centering
\vspace{1.5cm}
\includegraphics[angle=90,width=15cm]{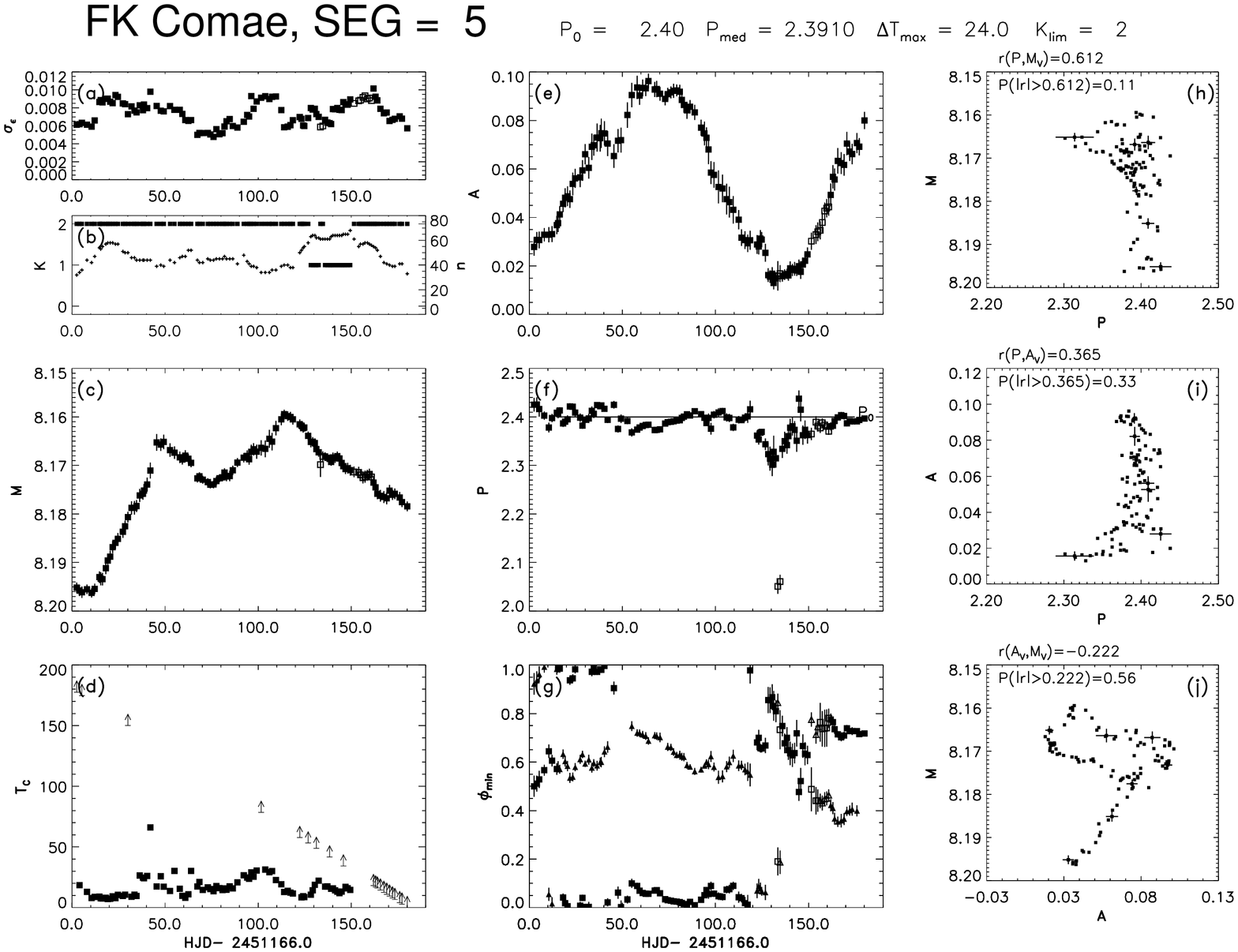}
\caption{The CPS analysis of segment SEG5 of the 
Amadeus photometry of \object{FK Com}. The phases in panel (g) where 
calculated using the period $P_\mathrm{med}=2\fd3910$.}
\label{seg5}
\end{figure*}

\section{Results}
\label{results}

The results from the CF analysis are shown in the fifteen
panels of Fig.~\ref{alldataCF}, wrapped with the carrier period
$P_0=2\fd40$
and assigning the phase $\phi=0$ to the first time point of each segment.
The phase information of the segments is therefore
not comparable. The dark-bright pattern is replicated as the phase-axis is
extended around $[0,1]$ to help the visualisation. The dark colours 
represent higher magnitude, i.e. lower temperatures.
The results of the fitting 
procedures are summarised in 
Table~\ref{CFres}, where
$R^2_1$ and $R^2_2$ are the coefficients of determination (Eq. \ref{detcoef}) 
before and after the
removal of the $3\sigma$ outliers, respectively. $\Delta N$ is the number
of removed data points. As an example, the fits for two segments are shown in 
Fig. \ref{seg4-5cffit}.

\begin{table}
\caption{Summary of CF fit results for each segment.}
\begin{center}
\begin{tabular}{llll}
\hline \hline
Segment &$R^2_1$ [$\%$]  &$\Delta N$ &$R^2_2$ [$\%$]\\ \hline
SEG1    &-      &-   &-    \\
SEG2    &95.3   &-   &-    \\
SEG3    &96.4   &-   &-    \\
SEG4    &62.9   &4   &83.0 \\
SEG5    &83.6   &6   &86.7 \\
SEG6    &95.0   &-   &-    \\
SEG7    &97.7   &-   &-    \\
SEG8    &93.0   &-   &-    \\
SEG9    &88.0   &6   &90.7 \\
SEG10   &78.5   &6   &89.9 \\
SEG11   &95.7   &-   &-    \\
SEG12   &97.1   &-   &-    \\
SEG13   &98.0   &-   &-    \\
SEG14   &93.1   &-   &-    \\
SEG15   &89.1   &2   &91.0 \\
SEG16   &98.5   &-   &-    \\
\hline
\end{tabular}
\label{CFres}
\end{center}
\end{table}

The full CPS results of segments SEG4 and SEG5 are shown
in Figs. \ref{seg4} and \ref{seg5}. The values of the a priori period 
estimate $P_0$, the median of all reliable periods
$P_\mathrm{med}$, the limiting modelling order $K_\mathrm{lim}$ 
and the maximum length of the dataset $\Delta T_{\max}$ are given at the 
top of the plots. The panels show: 

\begin{description}
\item (a) standard deviation of residuals $\sigma_\epsilon(\tau)$, $\tau$ being 
the mean epoch of each data set;
\item(b) modelling order $K(\tau)$ (squares, units on the left y-axis); and
number of observations per dataset $n$ (crosses, units on the
right y-axis);
\item (c) mean V-magnitude $M(\tau)$;
\item (d) time scale of change $T_C(\tau)$;
\item (e) amplitude $A(\tau)$;
\item (f) period $P(\tau)$;
\item (g) primary (squares) and secondary (triangles) minimum
phases $\phi_{\min,1}(\tau)$ and $\phi_{\min,2}(\tau)$. These
phases are calculated using the median period $P_\mathrm{med}$ of the segment;
\item (h) $M(\tau)$ versus $P(\tau)$;
\item (i) $A(\tau)$ versus $P(\tau)$;
\item (j) $M(\tau)$ versus $A(\tau)$. 
\end{description}

In the subplots (a), (c) and (e)--(g), the reliable parameter estimates
are indicated by filled symbols and unreliable ones by open symbols. The 
reliability is tested  as described by \cite{lehtinen2011}. In subplot (d), the
upward pointing arrows signify that 
the data fit the model, within the statistical limits explained by
\cite{lehtinen2011}, from this point on throughout the segment.

In the correlation plots (h)--(j), the error bars have been drawn only for 
the independent parameter estimates. The linear Pearson correlation
coefficients $r_0$ for the independent datasets, as well as an estimate of 
the probabilities $P(\vert r \vert > r_0)$, are given. 

\begin{figure*}
\centering
\vspace{-2.8cm}

\includegraphics[angle=90,width=14cm]{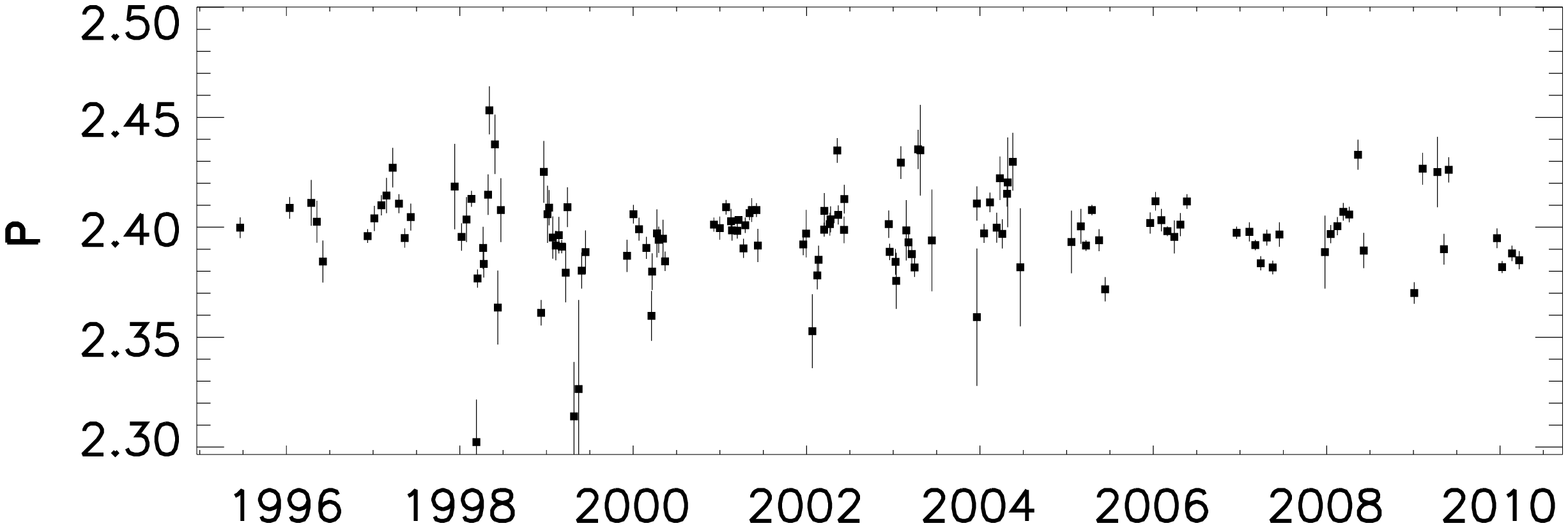}

\vspace{-5cm}
\includegraphics[angle=90,width=14cm]{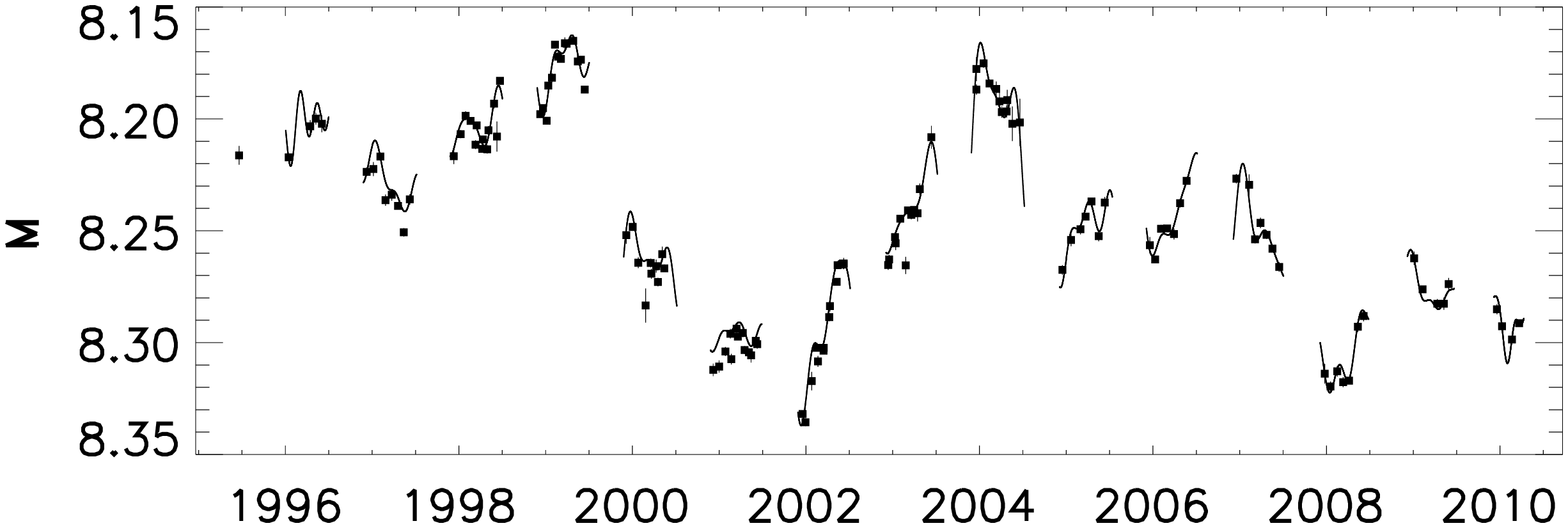}

\vspace{-5cm}
\includegraphics[angle=90,width=14cm]{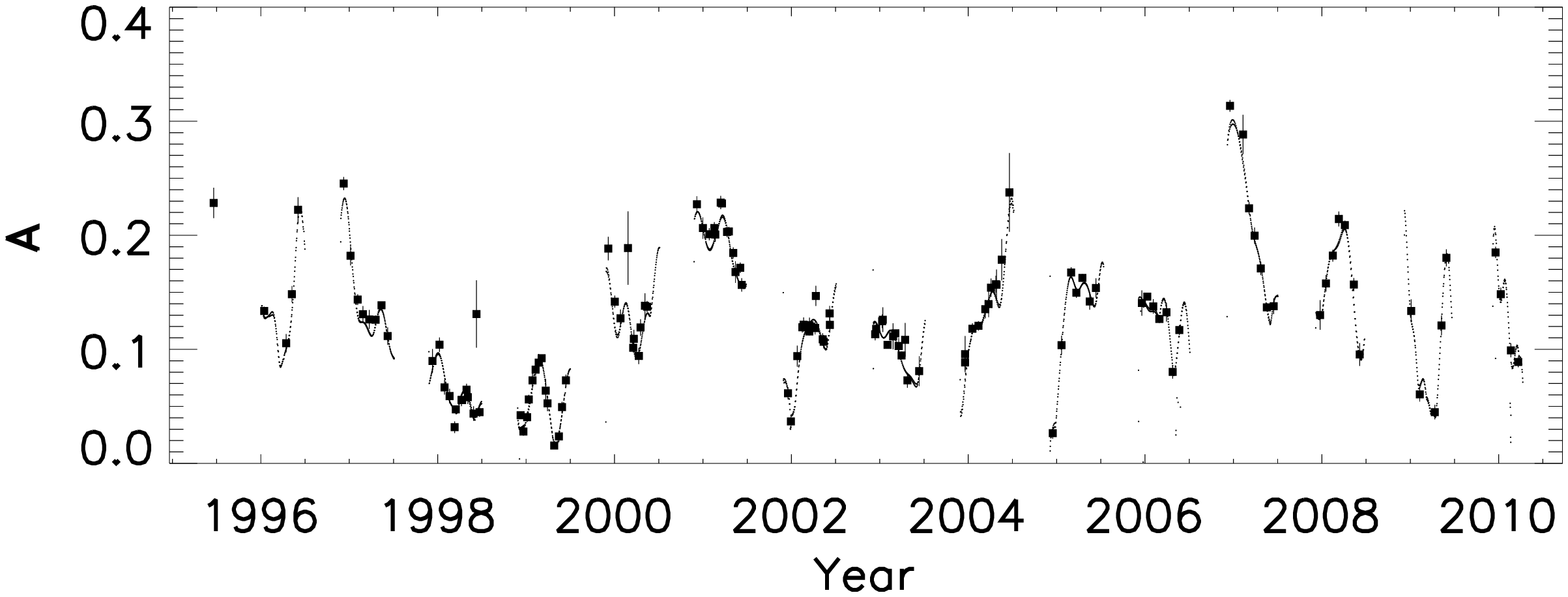}

\vspace{-2cm}
\caption{Independent period, mean magnitude and amplitude estimates from the CPS analysis(squares with error bars). The mean magnitude and
amplitude are plotted with the results from the CF analyses (dotted line).}
\label{allpma}
\end{figure*}

\begin{figure*}
\centering
\vspace{-2cm}
\includegraphics[angle=90,width=10cm]{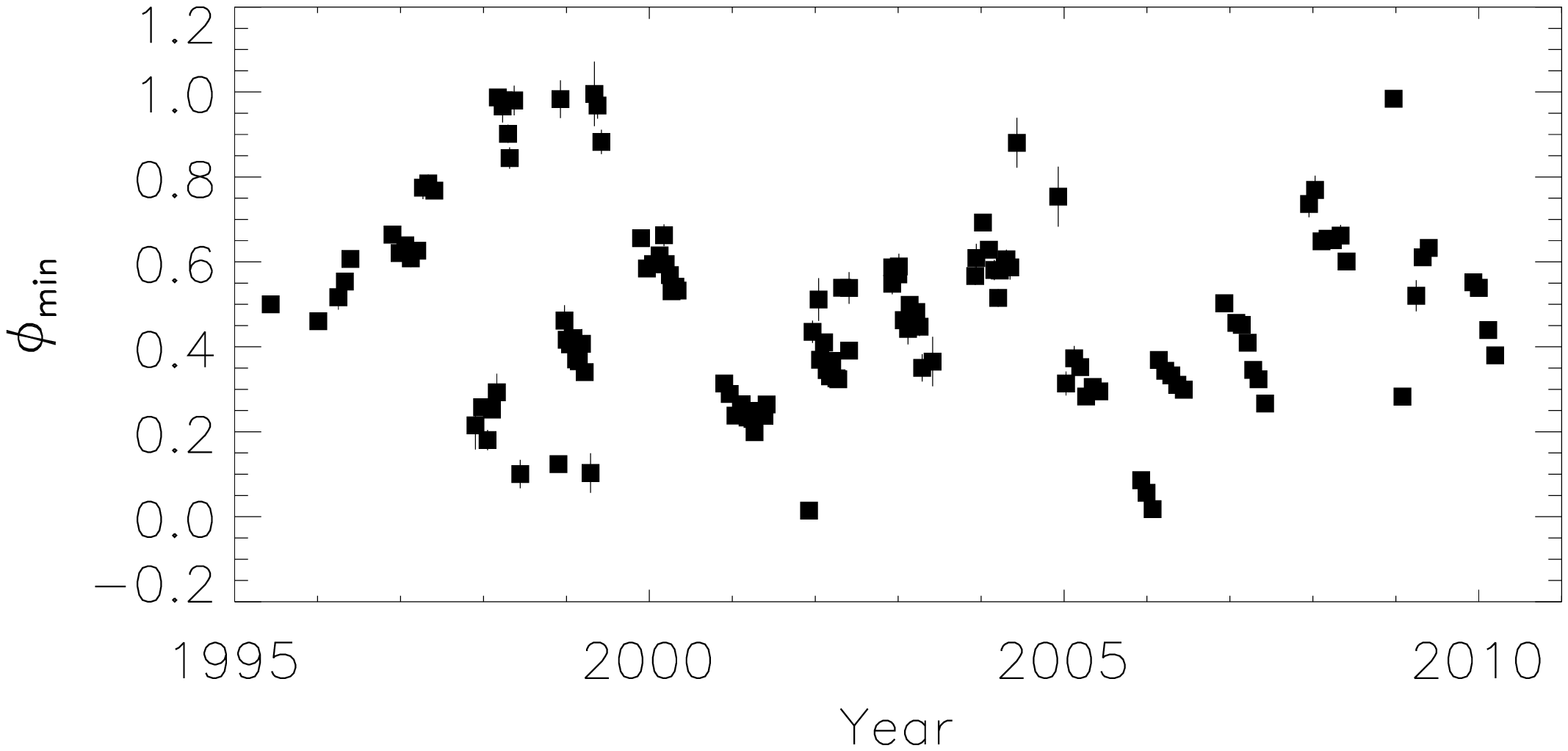}
\includegraphics[angle=90,width=7cm]{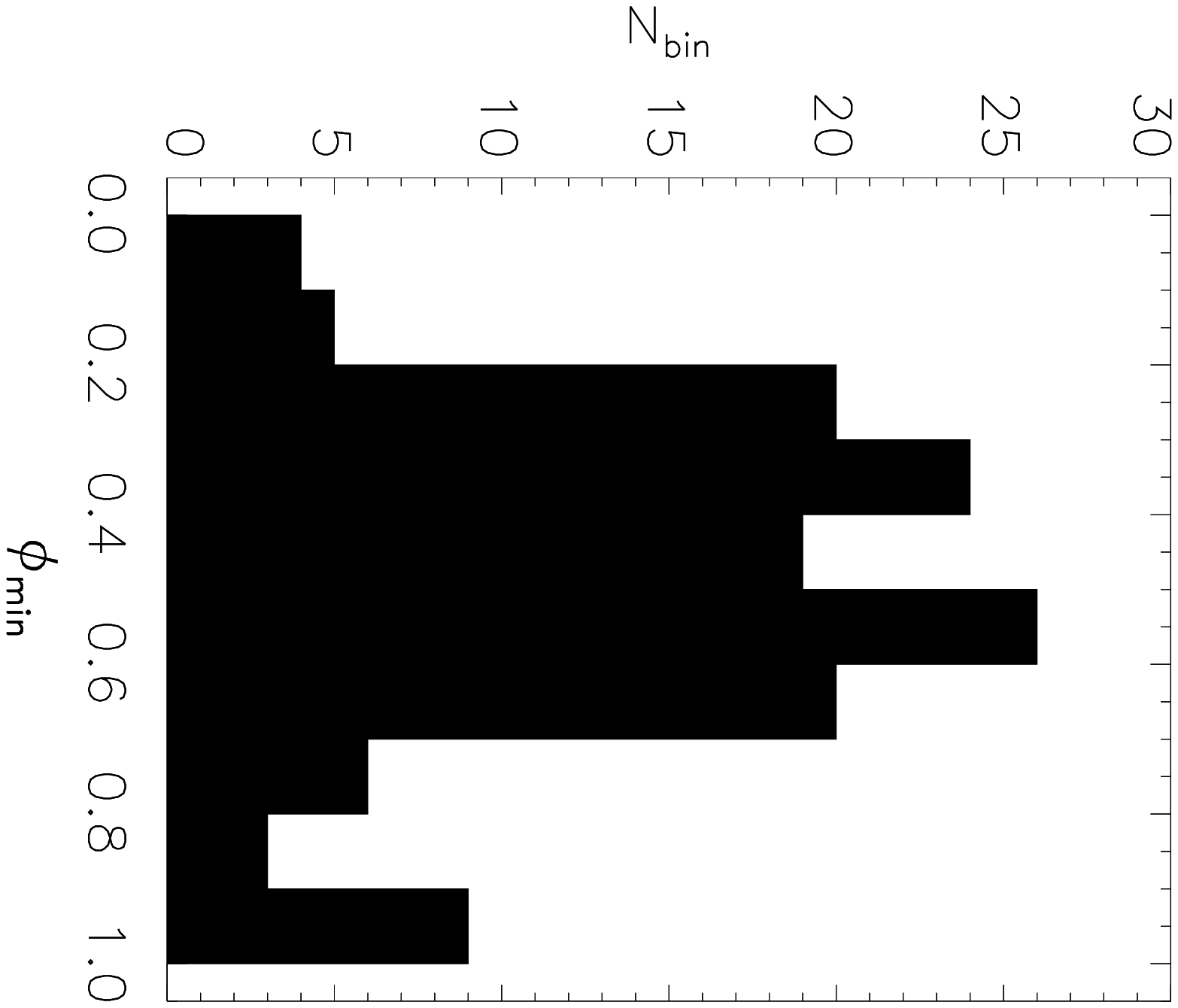}
\caption{The 
phase distribution of the
photometric minima 
folded with the period $P_\mathrm{al}=2.4012$. The 
epoch of the first minimum $t_{\mathrm{min}_1} = 24449877.321$ (HJD) 
corresponds to phase $\phi=0.5$.}
\label{min}
\end{figure*}

All independent CPS period estimates are shown in Fig. \ref{allpma}.
The mean period and its standard deviation were
$P_\mathrm{w}\pm\Delta P_\mathrm{w}\approx  2\fd3975 \pm 0\fd0123$. 
There is no doubt, 
that the photometric period is varying. The probable reason is differential
rotation. The variable period poses problems on what to use as a standard 
period for plots. We use the ``old'' ephemeris

\begin{equation}
\mathrm{HJD}_{\min} = 2439252.895 + 2.4002466E
\label{oldeph}
\end{equation}

\noindent
derived by \cite{jetsu1993} whenever we are combining results from the CF, CPS 
or Doppler imaging analysis. In the rest of the plots, we use either the 
carrier period (Fig. \ref{alldataCF}), the CPS median period for each 
segments (Figs. \ref{seg4} and \ref{seg5}) or the best active longitude 
period (Fig. \ref{min}).

\begin{figure*}
\centering
\includegraphics[trim=0.9cm 2.3cm 0cm 0cm, clip=true, angle=0,width=9cm]
{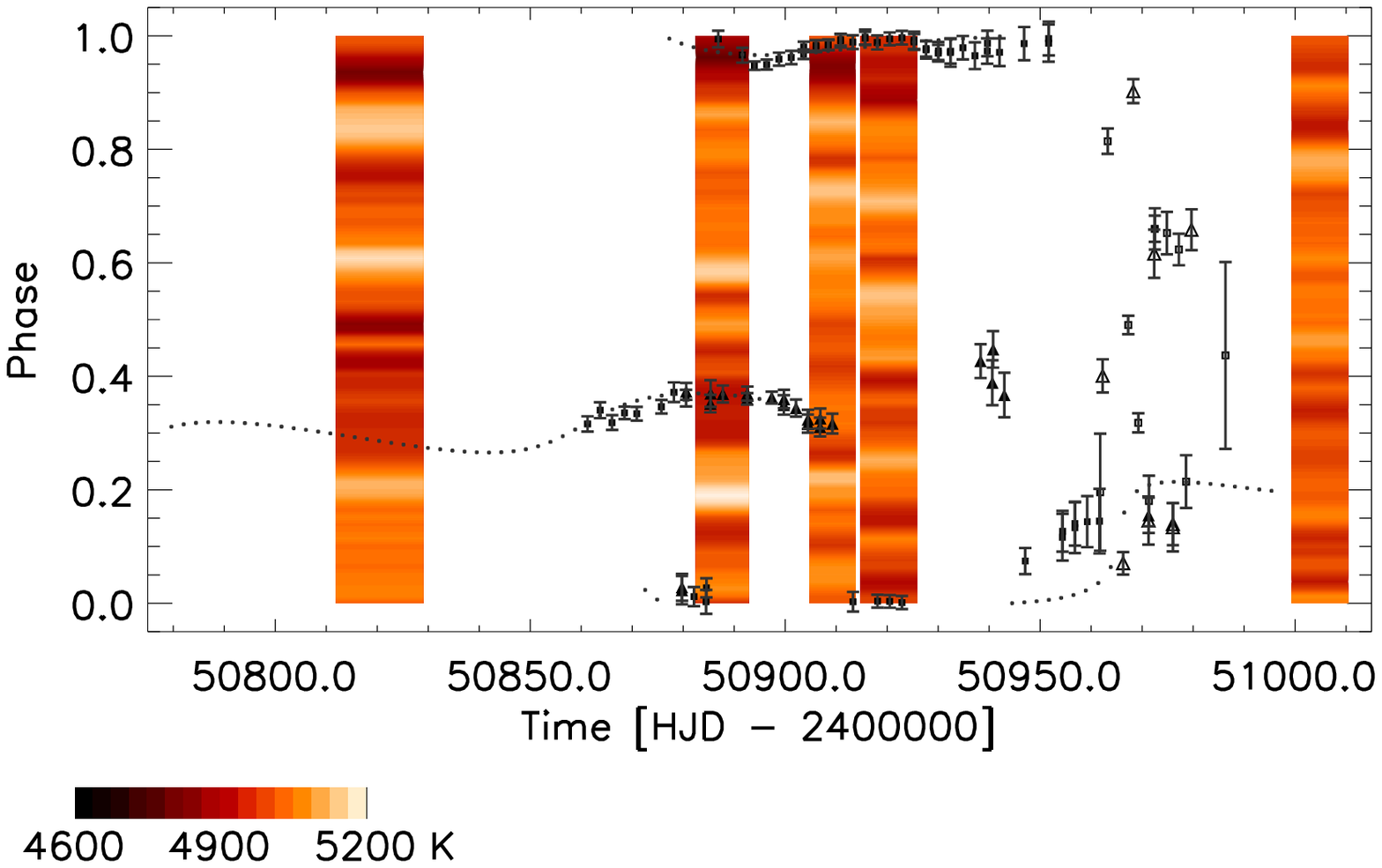}
\includegraphics[trim=0.9cm 2.3cm 0cm 0cm, clip=true, angle=0,width=9cm]
{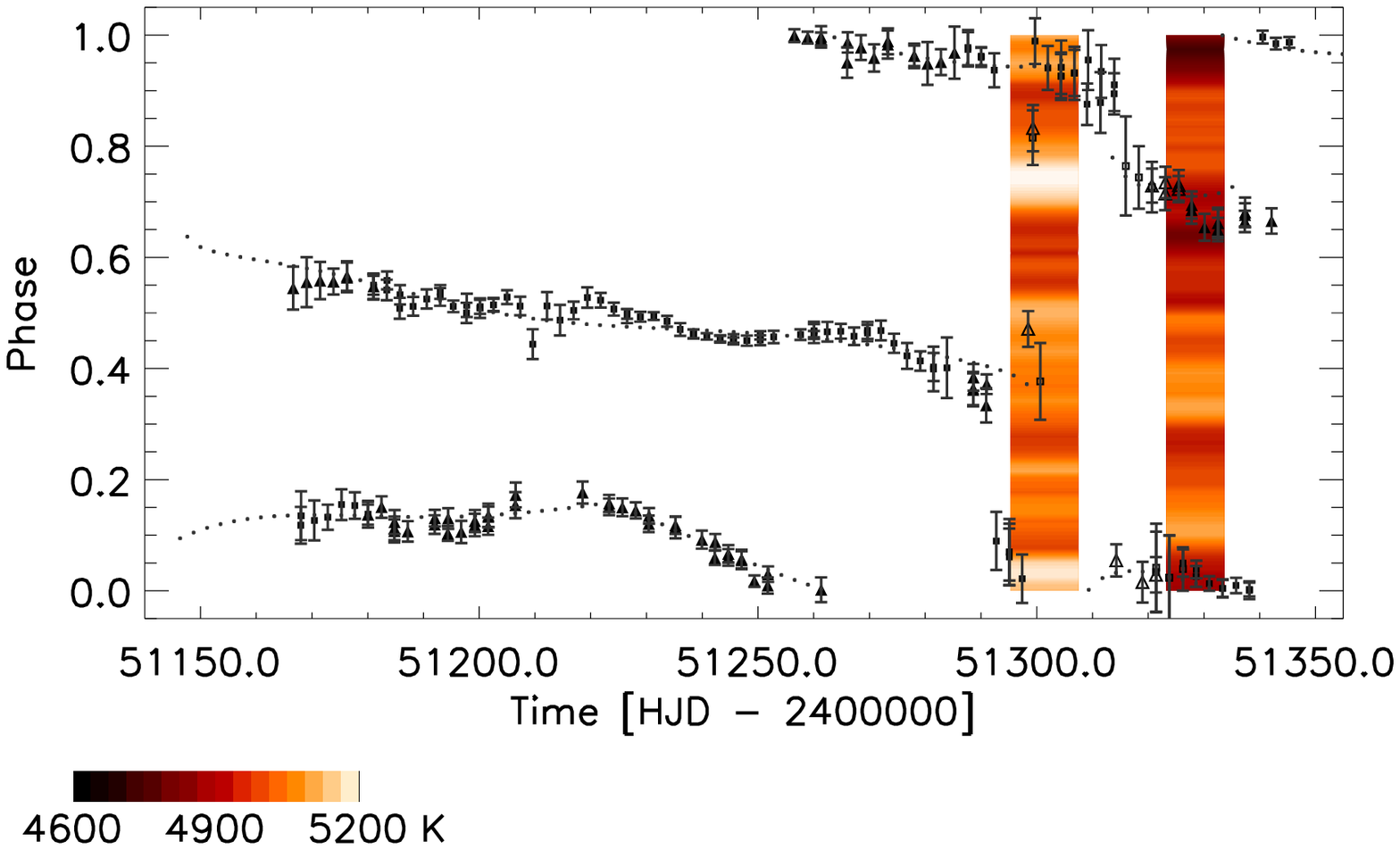}
\includegraphics[trim=0.9cm 2.3cm 0cm 0cm, clip=true, angle=0,width=9cm]
{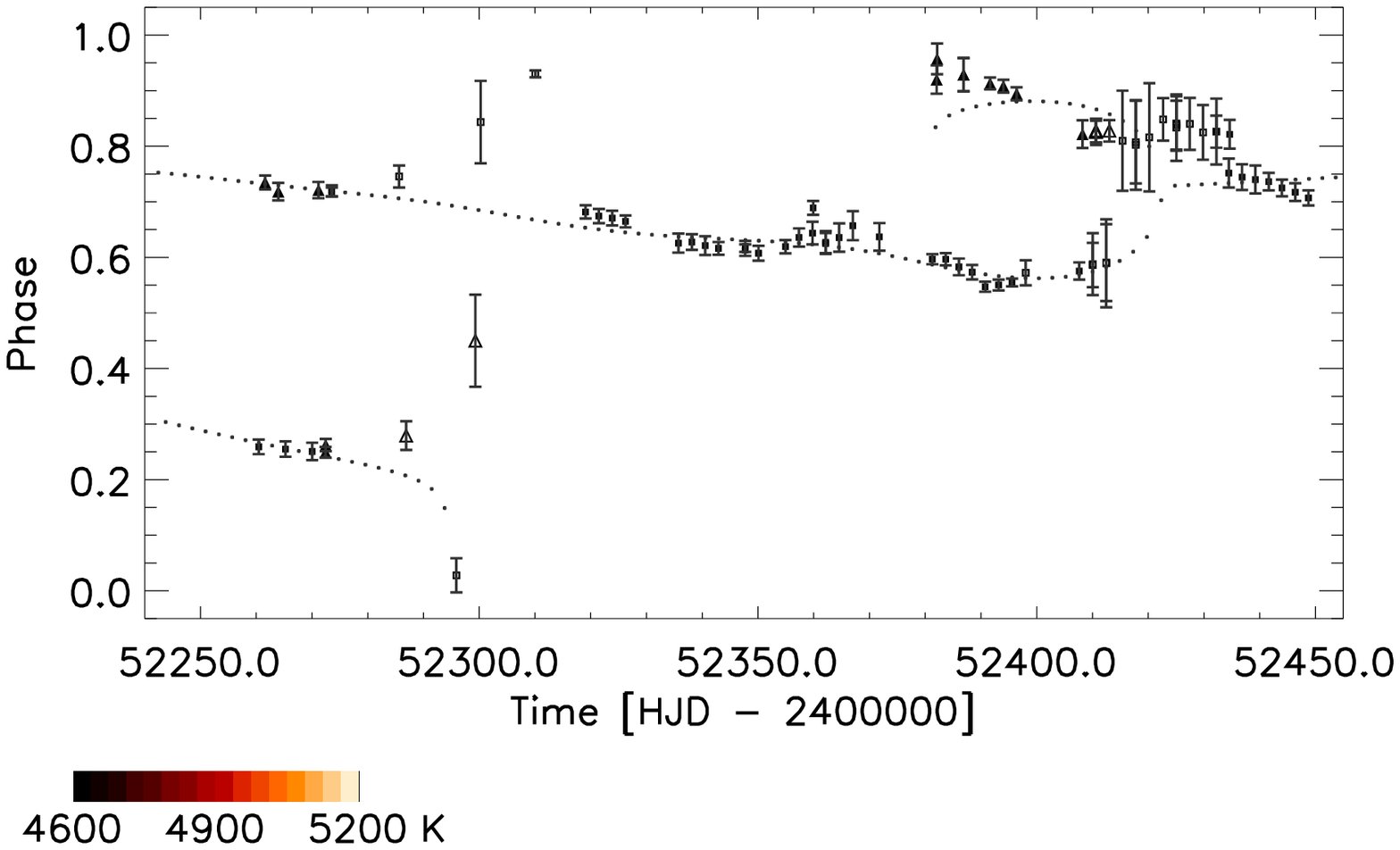}
\includegraphics[trim=0.9cm 2.3cm 0cm 0cm, clip=true, angle=0,width=9cm]
{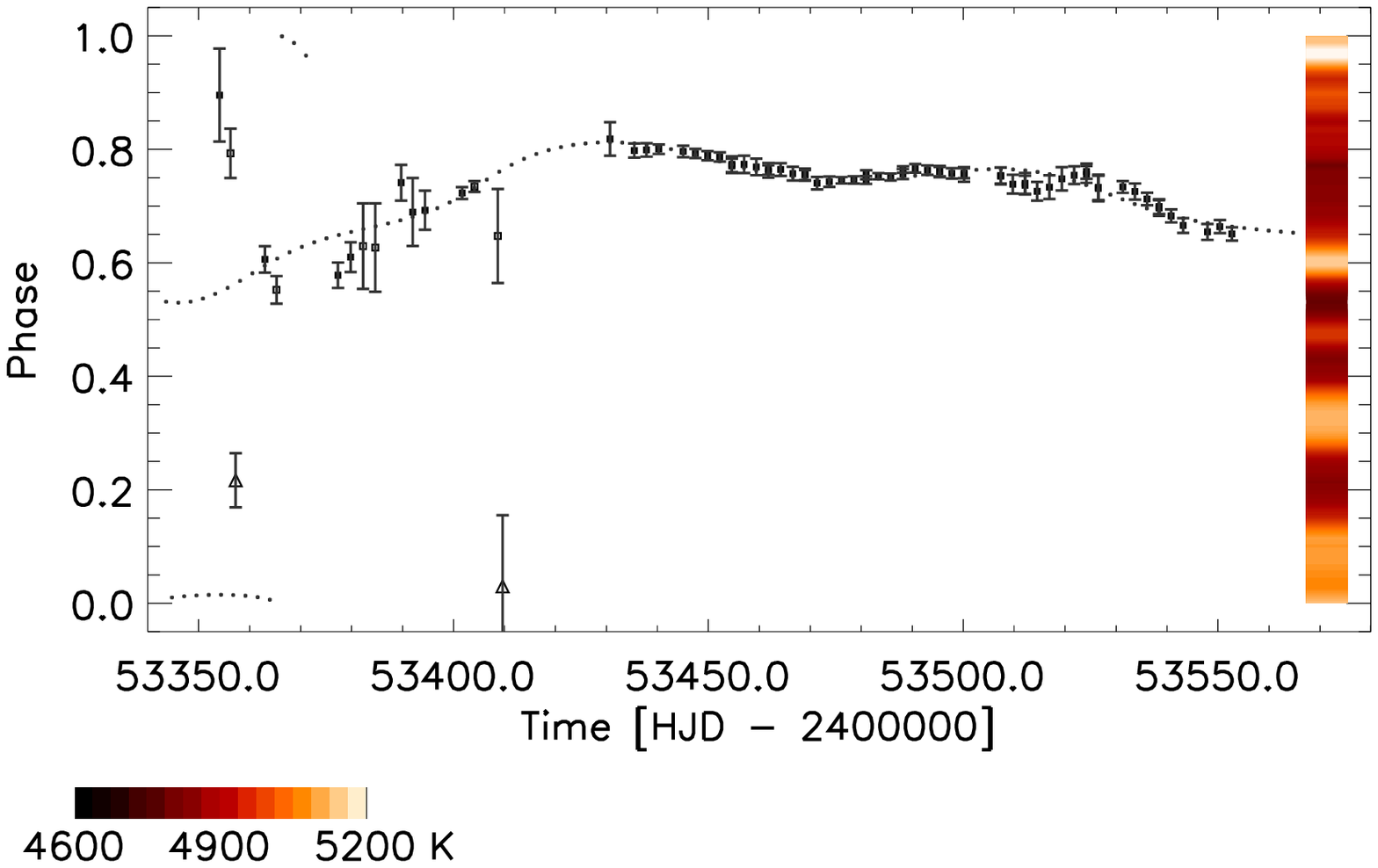}
\includegraphics[trim=0.9cm 0.1cm 0cm 0cm, clip=true, angle=0,width=9cm]
{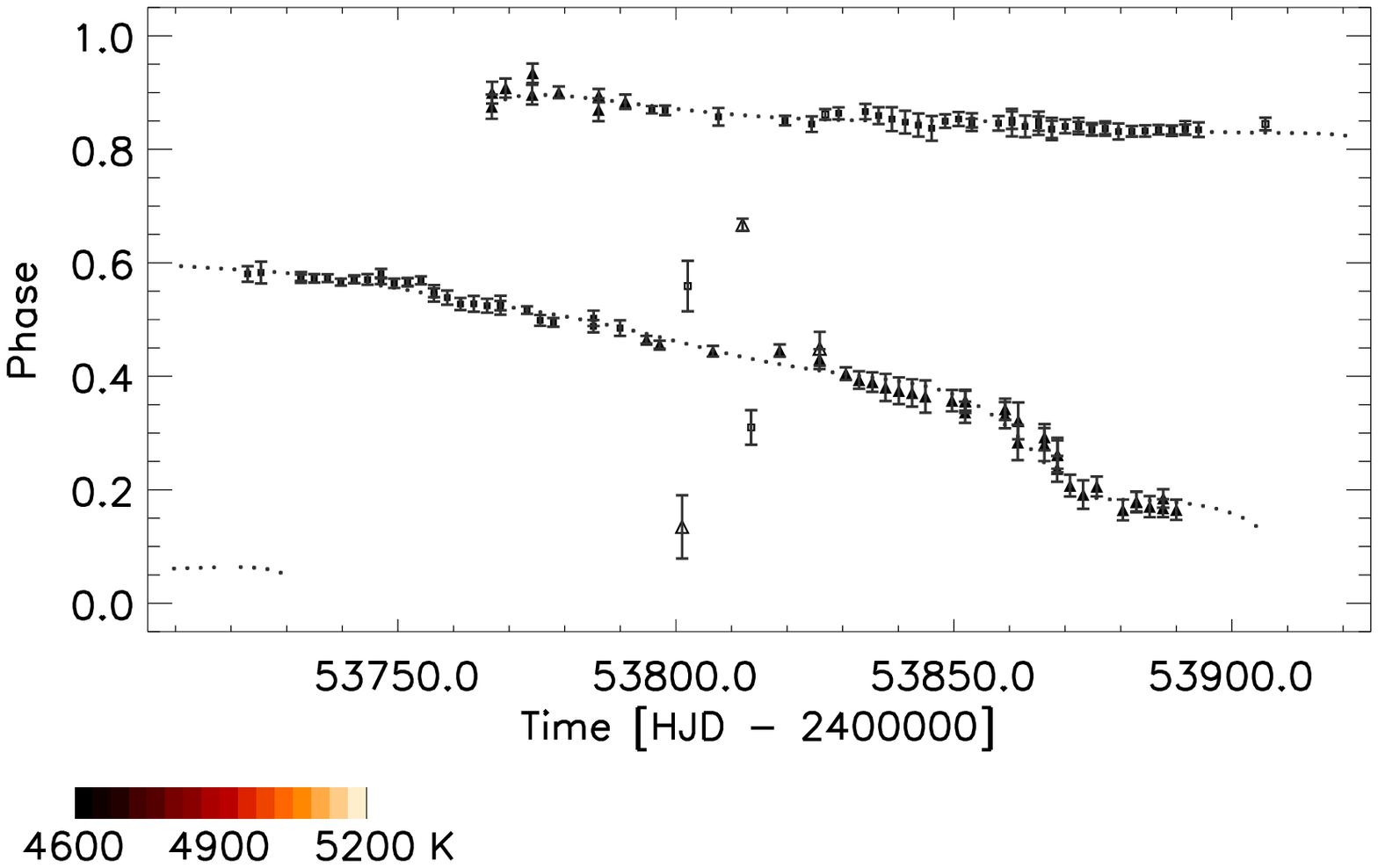}
\vspace{-0.5cm}
\includegraphics[trim=0.9cm 0.1cm 0cm 0cm, clip=true, angle=0,width=9cm]
{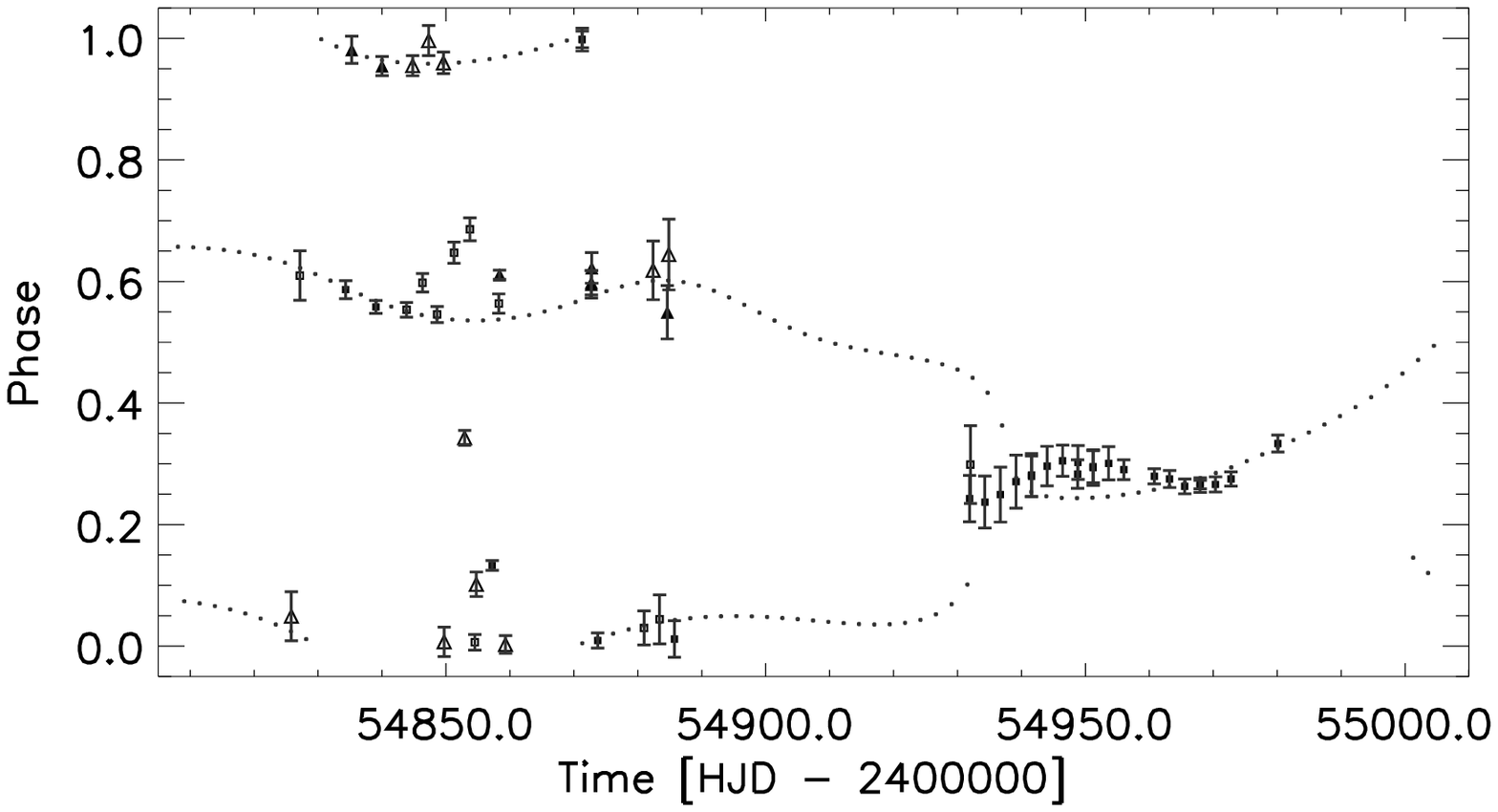}

\caption{The photometric minima from the CF and CPS analysis of segments from
left to right SEG4, SEG5, SEG8, SEG11, SEG12 and SEG15 plotted with 
longitudinal average 
slices of simultaneous Doppler images, when available. The dotted lines 
represent the primary and secondary minima from the 
CF analysis. The squares (primary) and triangles (secondary) show the
CPS minima. 
The phases were calculated using the ephemeris $ \mathrm{HJD_{min}} = 
2439252.895 + 2.4002466E$.}
\label{seg4zoom}
\end{figure*}

The independent CPS estimates of the mean magnitudes and light curve amplitudes
are plotted together with the corresponding results from the CF analysis in
Fig \ref{allpma}. In Fig. \ref{seg4zoom}
we show the CF and CPS photometric minima 
for some interesting segments,
together with results from Doppler imaging when available. 
In the latter cases we plot 
latitudinally averaged slices of Doppler images 
calculated by H. Korhonen \citep{ayres06,korhonen2007,korhonen2009,
korhonen2009b}. These slices were calculated
from the original rectangular DI maps by weighting each surface element with 
its size. For segment SEG5 we also plot all reliable CPS light-curves in Fig. 
\ref{seg5allc}.

We can see that there is no discrepancy between the CF and CPS
methods, since all differences can be explained by the different
approaches. The results from the CF method look like a smooth fit to
the more noisy CPS results. 

The mean of $T_C(\tau)$ is 32 d and the minimum 7 d. This is
both surprising and worrying. In previous analysis of main sequence stars
$T_C$ has been of the same order as the convective turnover time 
$\tau_c$ \citep{lehtinen2011,lehtinen2012}. It is hard to determine $\tau_c$ 
for \object{FK Com} since it is an exceptional giant star with uncertain 
parameters. However, with an estimated radius of $R \sim 10 R_{\sun}$, one 
would expect $\tau_c$ to be different than for main sequence stars. Thus one 
would expect also $T_C(\tau)$ to differ.
The worrying part is that about 30 \% of the $T_C(\tau)$ are smaller
than the $\Delta T_{\max}=24$ d used as the time 
window for the CPS-analysis. Tests with reducing $\Delta T_{\max}$ showed, 
however, that there was no significant
change in the result, except that the reduced number of points in each CPS-set
increased the errors. A low $T_C(\tau)$ indicates that rapid changes
occur in the spot configuration. It should be emphasised that
$T_C(\tau)$ of course depends on $K$: The higher the order of the
model, the more sensitive it is to changes. Furthermore, the analysis
is based on statistics, which explains why some low amplitude models may 
succeed in fulfilling the whole data, while subsequent models may be
quickly rejected (see e.g. the three ``arrows'' in the upper left corner of 
panel (d) in Fig. \ref{seg5}). Details on the estimation of $T_C(\tau)$ are 
found in \cite{lehtinen2011}.

\subsection{Active longitudes}

The overall result is that in about half of the segments only one
active longitude can be detected. Two active longitudes usually seem
to be present before and after longitudinal shifts in the activity,
but there are also cases, e.g. SEG16, of two simultaneous active
longitudes without indications of shifts. In some panels of
Fig. \ref{alldataCF} the pattern runs slanted upwards, in a couple it
remains nearly horizontal, in a few panels it is slanted downwards, or
in some segments disrupted in some way or another. The rising
(e.g. segments SEG2 and SEG3) or falling (e.g. SEG6, SEG9 and SEG13)
trends are indications of the carrier period not being optimal for the
segment, i.e. rising trends could be corrected with increasing the
carrier period somewhat, and correspondingly falling trends by
decreasing it. As both types of trends are present in the segments,
however, a global CF analysis of all the segments together would, in
any case, give a carrier period very close to the one already
adopted. The trends are relatively short-lived, as they are visible
only in two, maximally three consecutive segments, i.e. last roughly
one year. 

A drift longer by an order of magnitude has been reported
in the RS CVn binary II Peg \citep{lindborg2011,hackman2011} during
the years 1994--2001. It was interpreted as a possible azimuthal
dynamo wave arising from the properties of the non-axisymmetric dynamo
solution. During these years, the spot-generating structure was
observed to rotate with a shorter, but constant period, i.e. forming 
a more or less rigidly rotating entity.

The short-lived trends are likely to be related to
rotational non-uniformities either on the surface or at larger depth
that the spots may be anchored to. This can also be seen in the varying
photometric period retrieved with the CPS method (Fig. \ref{allpma}).
With the seasonal gaps in the
data, these trends make it hard to
visually follow any long-lived active longitudes in FK Com (see e.g. Fig.
\ref{pht}). 

However, the Kuiper periodogram analysis of the independent photometric minima 
from the CPS-analysis gave the best period $P_\mathrm{al}\approx 2\fd401151 \pm 
0\fd000092$ with a significance level of $Q \approx 5.3 \cdot 10^{-11}$.
The phases of the CPS minima, folded with this period, 
are displayed in Fig. \ref{min}. 
We also applied the Kuiper test on all photometric minima retrieved by the 
CPS method, as well as all minima and maxima from the CF-analysis. 
For these tests extremely low $Q$-values were derived, but since the 
measurements cannot be seen as independent, this has no 
statistical relevance. All the CPS-minima yielded the period  
$P_\mathrm{al}\approx  2\fd401173 \pm 0\fd000015$ and the CF-minima gave 
the result  $P_\mathrm{al}\approx 2\fd4011668 \pm 0\fd0000091$. However, the most
significant period for the CF-maxima was $P_\mathrm{max}\approx 2\fd405497 \pm 
0\fd000042$. 

Thus, the analysis of the CF-minima gave practically
the same $P_\mathrm{al}$ as for the 
CPS-minima. We note that $P_\mathrm{al}$ is slightly longer than the mean 
photometric period $P_\mathrm{w}$.
The value of $P_\mathrm{al}$ should describe the possible period 
of a magnetic structure within the star, while $P_\mathrm{w}$ may reflect
the rotation at 
the surface. This would imply, that there is a magnetic structure rotating
slightly slower than the surface of the star. The best period for the 
CF-maxima $P_\mathrm{max}$ was again slightly longer than $P_\mathrm{al}$. Thus,
this periodicity may describe something else than the mere absence of 
spots, i.e. possibly bright surface features.

\subsection{Flip-flop like events}
\label{ff}

Switching active longitudes were detected in six of the 15 analysed
segments, namely SEG4, SEG5, SEG8, SEG11, SEG12 and SEG15. These events are
summarised in Table \ref{flipflops} and plotted in Fig. \ref{seg4zoom} 
together with longitudinal slices of Doppler images. In many cases these 
events were not ``proper'' flip-flops. Either the phase jump was 
considerably less than 0.5, or the event involved more spot evolution, than 
just a shift from one active longitude to another one. In order to
make a distinction between these events we use the definitions for a
flip-flop suggested by \cite{kajat2013}:

\begin{itemize}
\item the region of main activity shifts about 180 degrees from the
old active longitude and then stays on the new active longitude.\\
or
\item the primary and secondary minima are first separated by
about 180 degrees, after which the secondary minimum evolves into a
long-lived primary minimum, and vice versa.
\end{itemize}

In SEG4 ($\mathrm{HJD}_0=2450778.043$), two phase changes can be detected from 
the CF analysis (upper left panel of
Fig. \ref{seg4zoom}). During the first phase disruption, the 
primary minimum segregates into a secondary minimum roughly at
$\mathrm{HJD} \approx 2450870$. The two minima have a phase separation
of $\Delta\phi\approx0.4$, which is close enough to 0.5 to call the event
a flip-flop. At around $\mathrm{HJD} \approx 2450880$, the secondary
minimum has become the main minimum. The CPS analysis shows, that the
two parallel minima persist for approximately 35 days. In this respect, the 
phase change is, of course, very rapid, as it occurs during roughly a month. 
Towards the end of this segment, another phase jump is observed. 
The phase change is only $\Delta\phi\approx0.2$ and based on the CPS analysis,
it occurs gradually during $\sim$ 20 days.

\begin{figure*}
\centering
\includegraphics[angle=90, width=17cm]{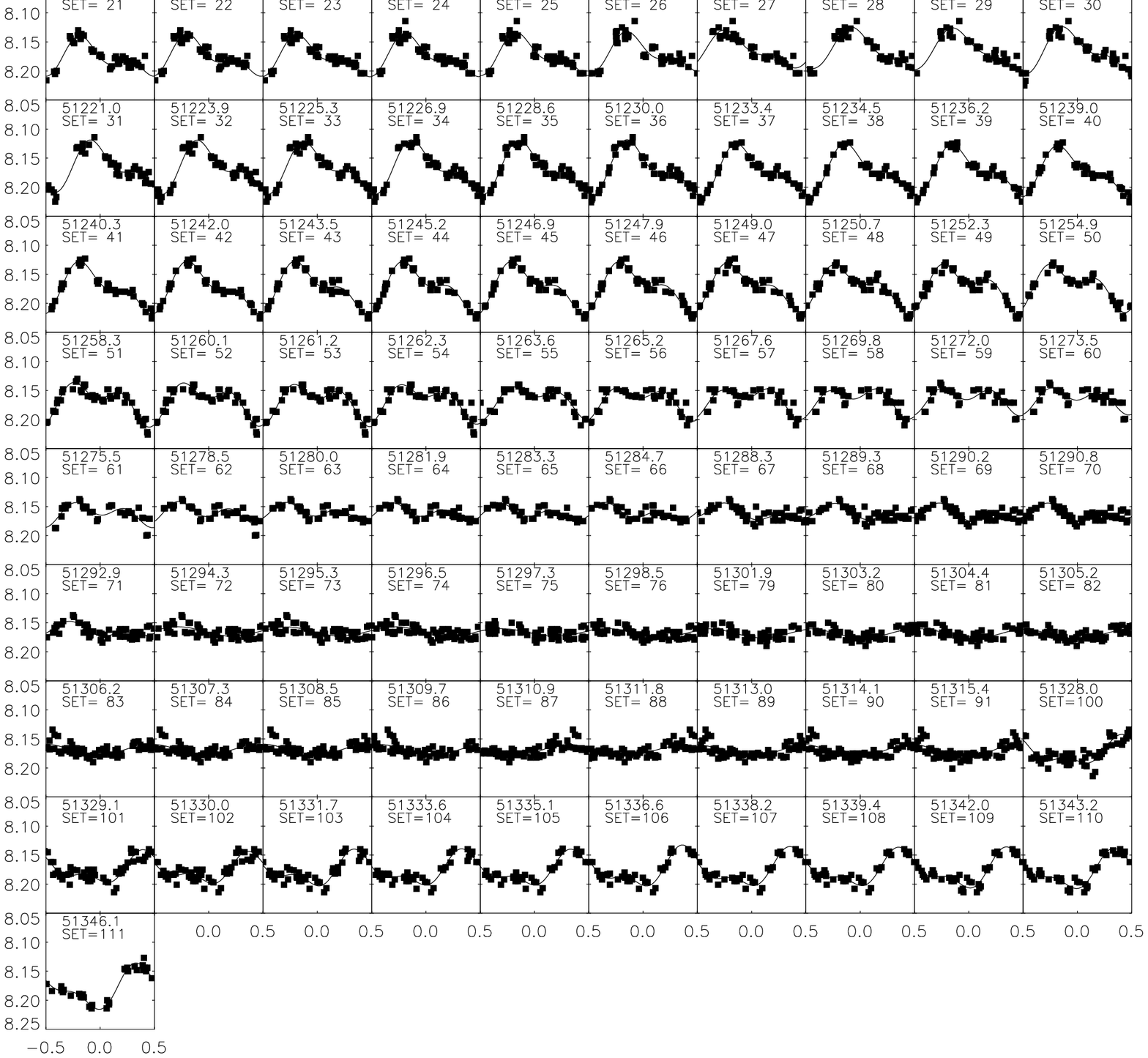}
\caption{All reliable light curve fits for SEG5. The phases
were calculated with the period derived for each set.}
\label{seg5allc}
\end{figure*}

In the beginning of SEG5 ($\mathrm{HJD}_0 =2451144.0394$) two active 
longitudes reside 
at opposite sides of the star. Both the CF- and 
CPS-method results show that the regions are not completely stable, but move
in phase (upper right panel of Fig. \ref{seg4zoom}).
This is also evident from the evolution of the 
light curve during this segment (Fig. \ref{seg5allc}).
At $\mathrm{HJD} \approx 2451170$ there is a flip-flop switch between the spot 
regions and the former secondary minimum becomes the new primary minimum.
The CPS-analysis also reveals that at $\mathrm{HJD} \approx 2451280$
one of the spot regions
starts drifting towards the other one and seems to pass it 
around $\mathrm{HJD} \approx 2451300$ (panel (g) in Fig. \ref{seg5}). The drift 
continues until $\mathrm{HJD} \approx 
2451330$. This could be explained by 
the fact that differential rotation will make 
surface features at different latitudes or with different anchor depths
move with different angular velocity.  
This drift is not apparent in the CF-analysis, as expected, because this method
is not supposed to register such fast changes. Instead we see a phase 
jump of $\Delta \phi \approx 0.3$ occurring after $\mathrm{HJD} \approx 
24515300$. From the plots of the individual 
light curves of the CPS-analysis we clearly see that the changes are actually
gradual, and not just an effect of ``interpolation'' over an 
abrupt change (Fig. \ref{seg5allc}).
Panels (h)--(j) of Fig. \ref{seg5} show that there are clear connections
between the light curve mean, amplitude and period. Especially interesting
is the initial correlation between $A$ and $M$, which is evolved to a loop
in the $(A,M)$ diagram. Correlations between neighbouring points 
in this kind of diagram are expected, since these are not independent 
measurements
\citep[see][]{lehtinen2011}. But the loop indicates that there really
is a more complex connection between $A$ and $M$.
However, no such connections can be seen in segment SEG4 (Fig. 
\ref{seg4}).

The next detectable phase shift occurs in SEG8 
(middle left panel of Fig. \ref{seg4zoom}),
during which the primary minimum suddenly drops in strength, while a secondary 
minimum roughly $\Delta \phi \approx 0.5$ apart gains in magnitude, and 
during a few tens of days, the activity becomes concentrated 
to the location of the former secondary minimum. 
Although this event fulfils the criterion 
for a flip-flop, the new primary active longitude does not appear
to become completely stable. Another minimum emerges at $\mathrm{HJD} \approx 
2452380$ nearby it, gains in magnitude, and finally the two minima seem to 
merge.

In the beginning of SEG11 (middle right panel of Fig. \ref{seg4zoom}),
quite an abrupt phase shift,
of nearly $\Delta \phi = 0.5$, can be seen. This event, however, is
very close to the beginning of the data set and 
it is not detectable from the CPS analysis, probably due to the low light 
curve amplitude.

In the beginning of SEG12 (lower left panel of Fig. \ref{seg4zoom}),
the primary minimum is 
located at $\phi=0.6$ roughly for the first 60 days of observations.
At $\mathrm{HJD} \approx 2453770$ a secondary minimum emerges
at $\phi=0.9$, and an exchange of activity levels occurs between the active 
longitudes. At the same time, the former primary minimum slowly drifts to 
$\phi=0.2$. This event takes over 100 days

In the beginning of segment SEG15 (lower right panel of Fig. \ref{seg4zoom}),
both the CF and 
CPS-analysis reveal two active regions separated by $\sim 0.5$ in phase. 
Shifts in the strengths of these seem to occur at 
$\mathrm{HJD} \approx 2454830$ and $\mathrm{HJD} \approx 2454870$.
The two active longitudes 
seem to merge and form a common minimum at $\mathrm{HJD} \approx 2454940$.

In conclusion, the CF and CPS analysis disclose complex phase behaviour in
six of the analysed segments, namely SEG4, SEG5, SEG8, SEG11, SEG12, and SEG15.
The shifts in the primary minima in these segments can be explained by:

\begin{itemize}
\item Flip-flops as defined earlier in this section (SEG4, SEG5, SEG8, SEG11, 
SEG15; denoted ``ff'' in Table \ref{flipflops})
\item Phase jumps of $\Delta \phi < 0.4$ (SEG4, SEG12; denoted ``phj'')
\item Two drifting active regions (SEG5, SEG12; denoted ``dr'')
\item Merging of two active regions (SEG8, SEG15; denoted ``mr'').
\end{itemize}

\noindent
The main reason for differing between phase jumps and flip-flops is that this
has consequences for the dynamo mode dominating the spot activity during a
specific period.
The merging of the active regions do not necessarily mean a 
physical merger, but rather that the two regions form a common minimum. As 
showed by \cite{lehtinen2011}, there is a minimum phase difference for spot 
regions under which these cannot be observed to cause separate minima.

\begin{table}
\caption{Summary of the flip-flop like events in \object{FK Com}: 
Time, duration and type of event.}
\begin{center}  
\begin{tabular}{lccrc}
\hline \hline
Segment &HJD  & year & $\Delta t$ [d] & Event type$^{\mathrm{a}}$ \\
\hline
SEG4    &2450870 & 1998.2 & 35 & ff \\
SEG4    &2450960 & 1998.4 & 20 & phj \\
SEG5    &2451170 & 1999.0 & 10 & ff \\
SEG5    &2451270 & 1999.3 & 60 & dr \\
SEG8    &2452270 & 2002.0 & 20 & ff \\
SEG8    &2452420 & 2002.4 & 40 & mr \\
SEG11   &2453360 & 2005.0 & 30 & ff \\
SEG12   &2453790 & 2006.1 & 30 & phj \\
SEG12   &2453790 & 2006.1 & 110 & dr \\
SEG15   &2454870 & 2009.1 & 30 & ff \\
SEG15   &2454940 & 2009.3 & 50 & mr \\
\hline
\end{tabular}
\begin{list}{}{}
\item[$^{\mathrm{a}}$] ff = flip-flop, phj = phase jump, dr = drifting
spot regions, mr = merging spot regions
\end{list}
\end{center}
\label{flipflops}
\end{table}

The five
flip-flops occur at times (in years) $t \approx$ 1998.2, 1999.0, 2002.0, 
2005.0 and 2009.1. Thus, the interval between these events is 
0.8--4.1 years and
no clear periodicity can be seen. Furthermore, the division between flip-flops
and phase jumps is not completely clear, since there is a ``grey zone'' of
$\Delta \phi \approx 0.4$ phase shifts. We cannot rule out that the flip-flops
and phase jumps play a role in the possible activity cycle, but the present 
data is not sufficient for any definite
conclusions.

\subsection{Differential rotation}

In all segments we see drifts of the active regions. These indicate that
the spots are not rotating with constant angular velocity.
From the independent CPS period estimates we got the weighted mean period 
$P_\mathrm{w}\pm\Delta P_\mathrm{w}\approx  2\fd3975 \pm 0\fd0123$. 
If the variations in the photometric period were caused by differential 
rotation, we could estimate the differential rotation coefficient with the 
parameter $Z= {6 \Delta P_\mathrm{w} \over   P_\mathrm{w}}$ 
\citep{jetsu_lq_1993}. 
For FK Com,
we got the 
value $Z \approx 0.0308 $, which would correspond to a differential rotation
of $\Delta \Omega > 4.6 \degr$/d.

An alternative way to estimate the differential rotation is to study the
drifts of active regions. During segment SEG5 the separation of the two
active regions changed from -0.5 to 0.2 in 60 days. This would correspond
to a differential rotation of $\Delta \Omega \approx 4.2 \degr$/d between
the two spot structures. This 
is, as it should be, less than the value derived from the variations in the rotation period, but still
more than twice the value derived by 
\cite{korhonen2007} assuming a solar differential rotation law (Eq. \ref{dr}). 

\begin{figure*}
\centering
\includegraphics[angle=0, width=15cm]{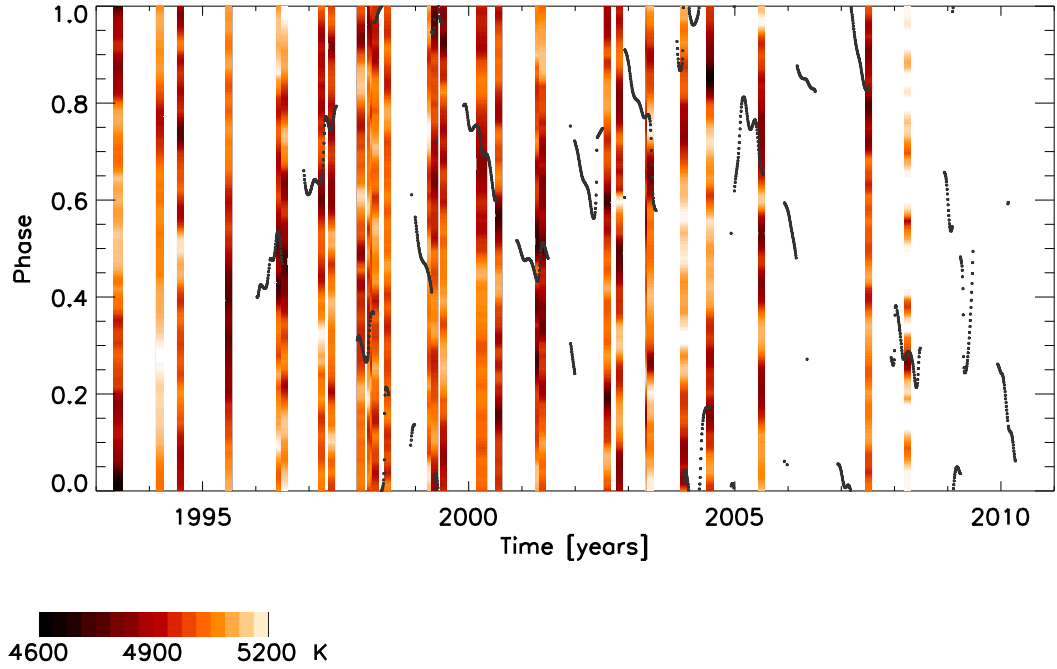}
\caption{Longitudinal slices of Doppler images and photometric minima from
the CF analysis. The phases were calculated using the same ephemeris as 
in Fig. \ref{seg4zoom}.}
\label{pht}
\end{figure*}

\subsection{Comparison with previous results}

Our analysis is partly based on the same data as used by \cite{olah2006} and 
covers in part the same time period as the photometric analysis of 
\cite{PanovDimitrov}
and the Doppler images by
Korhonen and collaborators
\citep{korhonen2007,korhonen2009,korhonen2009b}.
In Fig. \ref{seg4zoom} we already compared
our results with simultaneous Doppler imaging maps. In Fig. \ref{pht} we 
compare the primary minima from the CF analysis with Doppler images from the
years 1993--2008. This figure illustrates the rapid changes in the spot 
activity of \object{FK Com}. In general the consistency between the
CF analysis and Doppler images is satisfactory. However, one must take into
account that the photometric minimum is a result of the integrated effect 
of several spot regions, including the visibility effect. In calculating the 
longitudinal Doppler imaging stripes, visibility and limb darkening was not 
taken into account.

Flip-flops and phase jumps during the time period covered by the current 
analysis have been reported in several papers 
\citep{korhonen2002,olah2006,PanovDimitrov,korhonen2009}. 
Interestingly \cite{olah2006} also 
recognised the complex phase behaviour in SEG5. 
They conclude that a flip-flop occurred during this segment, 
similarly to \citet{PanovDimitrov} and \citet{korhonen2009}, 
while we detect both a flip-flop and drifting spot regions. 
\citet{PanovDimitrov} reported a flip-flop occurring during the early 1998, 
which is also seen in the current analysis. Furthermore, 
\citet{olah2006} detected a phase jump in SEG3, and \citet{korhonen2000} 
and \citet{PanovDimitrov} reported a flip-flop in this segment. 
In our analysis this would be the ``bump'' visible in Fig. \ref{alldataCF}. 
Similar small bumps can be seen in several of the segments, which is why we did
not focus on this case. 

Our analysis includes observations of $\sim$ 15 years, which is less than 
3 $\times$ the cycles of 5.2 and 5.8 years reported 
by \cite{olah2006} and \cite{PanovDimitrov}. The long term changes of the mean
magnitude in our analysis (middle panel in Fig. \ref{allpma}) shows signs of
possibly cyclic variations on a time scale of 6 years, but no clear conclusions
can be drawn from our analysis because of the limited time span.

\begin{table}
\caption{Rotation periods used in the CF- and CPS-analysis and their 
mathematical or physical interpretation.}
\begin{center}
\begin{tabular}{lll}
\hline \hline

Symbol \ & Meaning & Interpretation \\
\hline
$P_\mathrm{phot}$ & Photometric rotation  & TSPA period 
\citep{jetsu1993} \\
                & period & from long-term photometry \\
$P_0$ & First guess for the CF-  & Approximate mean photometric  \\
                & and CPS-analysis & period used as the carrier period \\
$P_\mathrm{med}$ & CPS analysis median  & The median of all reliable \\
                & period & periods within a CPS-segment \\
$P_\mathrm{al}$      & Period of minima from & May describe the rotation of \\
                & the Kuiper-method & magnetic structures feeding \\ 
                &                        & the surface with spots\\
$P_\mathrm{w}$       & Mean CPS-period        & The mean of all reliable \\
                &                        &  periods within a CPS-segment \\
$P_\mathrm{max}$     &  Kuiper period & May reflect the bright \\
                & of CF-maxima & surface features \\
\hline
\end{tabular}
\label{periods}
\end{center}
\end{table}

\section{Discussion}

Our analysis clearly shows, that the behaviour of FK Comae cannot
  be explained by a single rotation period. Instead, we find multiple
  periodicities, the 
explanation
of the periods being listed in
  Table~\ref{periods} in an attempt to remove any confusion about
  them. These periods are probably related to differential rotation as well
as the interplay between rotation and varying surface spots.

Our close study of six flip-flop like events clearly shows that we are not 
dealing with a singular phenomenon. The shifts of the active longitudes can 
be a result of both abrupt and gradual changes. In one case the apparent 
flip-flop can be explained by two spot regions moving with different angular 
velocity and even passing each other. 
Since FK Comae is
a late-type star with a convection zone, some amount of differential 
rotation, changing both as a function of depth and latitude, can be expected to 
be generated. 
The apparent flip-flop can thus be 
a consequence of differential rotation in two alternative ways:

\begin{itemize}
\item The spot latitudes change and the spot or spot groups drift with 
respect to each other because of surface differential rotation.

\item The anchor depth of the spot or spot groups change and the differences
in the angular velocity is caused by the depth dependent differential
rotation.

\end{itemize}

An alternative way to produce flip flops, also involving
  differential rotation, is the competition of a solar-like oscillatory
  axisymmetric dynamo mode with a steady non-axisymmetric mode of
  comparable strength \citep[see e.g.][]{korhoetel}. This mechanism,
  would result in regular phase changes, which may be hard to detect because
  of the lack of a clear reference rotation period.

Furthermore, the situation will be complicated by rapid spot evolution. At 
times there are significant changes in the light curves within $\sim 10$ days.
This time would probably be much shorter with light curves with
denser spacing and higher accuracy, e.g. satellite observations.
The changes in the light curves of \object{FK Com} cannot be explained just 
by differential
rotation of a steady spot model (see e.g. Fig. \ref{seg4-5cffit}). In 
reality we may thus be witnessing a combination of several effects: Rapid
spot evolution combined with differential rotation in both depth and 
latitude, spiced with a possible dynamo wave.

We find that the differential rotation in \object{FK Com} is at least about 
$\Delta \Omega \approx 4.6 \degr/$d, which would correspond to a 
differential
rotation coefficient of $k \approx 0.03$. This is roughly three times larger 
than the value suggested by \cite{korhonen2007} and about one sixth of the 
solar value. 
Our new value is in fair agreement with the observational and theoretical consensus \citep[see e.g.][]{kitcha1999} of differential rotation diminishing proportional to 
$\frac{\Delta \Omega}{\Omega} \simeq \Omega^{-n}$,
where $n \approx 0.8 - 0.9$. The rotation period of the Sun being
roughly an order of magnitude slower than that of FK Com, indicates
roughly 7
times weaker differential rotation for FK Com.
However, we cannot conclude that this is a measure of the
surface differential rotation, as it could also reflect 
geometric properties of the large-scale dynamo field,
as described by \cite{korhoetel2011}, 
and include a signal
from a possible azimuthal dynamo wave \citep[e.g.][]{krause1980,
lindborg2011}.

The finding of an active longitude period, which is slightly longer than the 
mean rotation period indicates the presence of an azimuthal dynamo wave. In the
case of \object{FK Com} this would rotate slower than the star itself,
as opposed to the RS CVn star \object{II Peg} \citep{lindborg2011,
hackman2011,hackman2012}.
Even though a clear long-term active longitude period can be detected, the
short term spot evolution seems fast and random. Thus, it is 
not as easy to follow active longitudes over gaps in the data as in the
case of e.g. \object{II Peg}. 
In this sense \object{FK Com} resembles some other single stars, e.g. 
\object{HD 116956} 
\citep{lehtinen2011} and \object{LQ Hya} \citep{lehtinen2012}.
In general, the active longitudes of close binary stars may be more regular 
due to the tidal effects.

We also note that there is a clear tendency of having two active structures 
with a phase difference of 
considerably less than 0.5. This tendency can also be seen in the distribution 
of the minimum phases folded by $P_{\mathrm{al}}$: The maximum in Fig. \ref{min} 
is divided into two peaks with a separation of roughly 0.2. The same was 
already noted in earlier studies. 
E.g. \cite{korhonen2009b}
reported a pair of
spots in 2008 with a phase difference of 0.25.
This could 
resemble the pattern of leading and trailing spot pairs seen in much smaller 
scale on the Sun.

Concerning the possible existence of an activity cycle our study does not
bring much 
conclusive
evidence. The mean magnitude shows variability which
is compatible with 
the cycle of 4.5--6.1 years found by \cite{olah2009}.
Naturally, when two active longitudes
coexist, the amplitude tends to be lower. However, we do not see any other 
connection between the switches in the active longitudes and the long-term
photometric variability. Neither can we detect any regularity in the
flip-flop like events of \object{FK Com}.

\begin{acknowledgements}
  The work of TH was financed by the research programme ``Active
  Suns'' at the University of Helsinki. HK acknowledges the support
  from the European Commission under the Marie Curie Intra-European
  Fellowship. Financial support from the Academy of Finland grants
  No. \ 218159 (MJM) and 141017 (JP) are gratefully acknowledged. 
The work of PK was supported by the Vilho, Yrj\"o and Kalle V\"ais\"al\"a 
Foundation.
\end{acknowledgements}

\bibliographystyle{aa}
\bibliography{fkcom_fin}



\end{document}